\documentclass[manuscript,authorversion,nonacm]{acmart}
\usepackage{hyperref}

\usepackage{amsmath,amsfonts}
\usepackage{graphicx,color} 
\usepackage{textcomp}
\usepackage{soul} 
\usepackage{booktabs} 
\usepackage{array}
\usepackage{multirow}
\usepackage{subfig}
\usepackage[inline]{enumitem} 
\PassOptionsToPackage{hyphens}{url} 
\usepackage{url}
\usepackage{xcolor}
\usepackage{color,colortbl}
\usepackage{tcolorbox}
\usepackage{makecell}
\usepackage[font=small,skip=0pt]{caption}
\newcommand{\quotes}[1]{``#1''} 

\newcommand{\autorefappendix}[1]{\hyperref[#1]{Appendix~\ref*{#1}}}  

\begin{document}

\title{Work From Home and Privacy Challenges: What Do Workers Face and What are They Doing About it?}

\author{Eman Alashwali}
\affiliation{%
  \institution{King Abdulaziz University (KAU) and King Abdullah University of Science and Technology (KAUST)}
  \country{Saudi Arabia}
  \authornote{Eman Alashwali was a Collaborating Visitor at CMU while working on this paper.}
  }
\email{ealashwali@kau.edu.sa}

\author{Joanne Peca}
\affiliation{%
  \institution{Carnegie Mellon University (CMU)}
  \country{United States}
  }
\email{jpeca@cmu.edu}

\author{Mandy Lanyon}
\affiliation{%
  \institution{Carnegie Mellon University (CMU)}
  \country{United States}
  }
\email{mandy@cmu.edu}

\author{Lorrie Faith Cranor}
\affiliation{%
  \institution{Carnegie Mellon University (CMU)}
  \country{United States}
  }
\email{lorrie@cmu.edu}

\begin{tcolorbox}
This document is the author's manuscript for a paper accepted at the Journal of Cybersecurity, 2025.
\end{tcolorbox}

\thispagestyle{plain}
\pagestyle{plain}

\begin{abstract}
The COVID-19 pandemic has reshaped the way people work, normalizing the practice of working from home. However, work from home (WFH) can cause a blurring of personal and professional boundaries, surfacing new privacy issues, especially when workers take work meetings from their homes. As WFH arrangements are now standard practice in many organizations, addressing the associated privacy concerns should be a key part of creating healthy work environments for workers. To this end, we conducted a scenario-based survey with 214 US-based workers who currently work from home regularly. Our results suggest that privacy invasions are commonly experienced while working from home and cause discomfort to many workers. However, only a minority said that the discomfort escalated to cause harm to them or others and that the harm was almost always minor and psychological. While scenarios that restrict worker autonomy (prohibit turning off camera or microphone) are the least experienced scenarios, they are associated with the highest reported discomfort. In addition, participants reported measures that violated or would violate their employer's autonomy-restricting rules to protect their privacy. We also find that conference tool settings that can prevent privacy invasions are not widely used compared to manual privacy-protective measures. Our findings provide a better understanding of the privacy challenges landscape that WFH workers face and how they address them, providing useful insights to organizations' policymakers and technology designers for areas of improvements, to provide healthier work environments to workers.
\end{abstract}

\maketitle
\section{Introduction}
\begin{figure}
\centering
\includegraphics[width=0.6\textwidth]{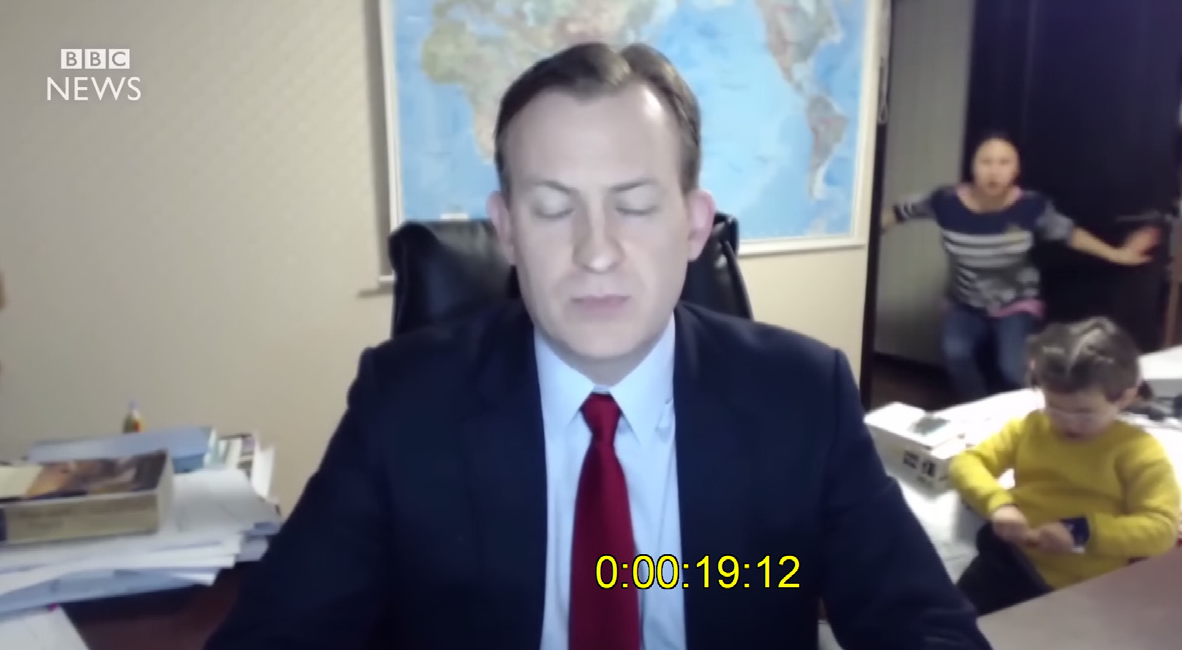}
\caption{Professor Robert Kelly, his child and wife invading a live streaming interview with the BBC channel \cite{bbc_news24}.}
\label{fig:bbc_dad}
\end{figure}

Work from home (WFH), is a flexible setting that allows workers to perform their work duties including collaboration with others such as meeting and teaching, remotely from their homes instead of a dedicated office space provided by employers. WFH often requires the use of information and communication technologies such as voice and video conferencing tools (e.g. Zoom \cite{zoom25} and Microsoft Teams \cite{teams25}). Before 2020, WFH was often provided as a benefit; however, in early 2020 with the advent of the global Coronavirus pandemic (COVID-19) \cite{who24}, WFH was no longer a benefit, but a mandate, despite the challenges it poses to some workers \cite{ford21,fukumura21,varner22}. WFH can be especially challenging for workers who live with children, lack appropriate space at home, or lack the skills needed to deal with new technologies. Some WFH challenges were exemplified in viral videos. For example, in 2017, Professor Robert Kelly's two young children invaded a live stream interview about South Korea's politics on the British Broadcasting Channel (BBC) and were subsequently dragged out of the room by his wife \cite{chappell20}, as illustrated in~\autoref{fig:bbc_dad}. 

After COVID-19, many organizations have changed their policies to allow WFH on an ongoing basis \cite{smite23,christian23,lufkin22}. According to a 2024 report by WFHResearch on US workers, the growth in WFH after COVID-19 is nearly equivalent to 40 years of growth before COVID-19 \cite{barrero24}. The same report shows that by early 2024, 13\% of full-time workers were working fully remotely, and 29\% were in a hybrid mode \cite{barrero24}. Data from the US Bureau of Labor Statistics show that in December 2024, 23.1\% of workers worked from home at least some of the time \cite{bls24}. 

With a significant fraction of the American workforce working from home all or part of the time, understanding the privacy concerns and behaviors that first arose out of an emergency, WFH arrangements should now be considered a key part of creating healthy work environments for workers. To do that, we must first understand the privacy concerns and behaviors of those whose \quotes{new normal} includes WFH on a regular basis. 

Our study aims to provide a detailed account of the privacy concerns workers face and the protective measures they take to prevent them. In particular, this study aims to answer the following main questions: \textit{How prevalent is experiencing privacy-invasive scenarios among WFH workers? To what extent does experiencing these scenarios cause discomfort? Does discomfort escalate to cause harm to workers or others? What type of harm do workers experience? What types of scenarios are associated with the highest reported discomfort? What protective measures do some workers take? Why do some workers choose not to take protective measures?}

To this end, we surveyed 214 US-based workers who currently work from home regularly about their experiences with privacy-invasive scenarios pertaining to: audio, video, data\footnote{For brevity, throughout the paper, we use the word \quotes{data} to denote \quotes{computer/browsing data.}}, and autonomy that can occur in a WFH setting. 

Our key findings can be summarized as follows:
\begin{enumerate}
    \item Privacy-invasive scenarios are commonly experienced and caused discomfort to most of our participants. Almost all participants (93.9\%) experienced at least one of the privacy-invasive scenarios, and more than half of the participants (65.4\%) felt uncomfortable during at least one of the scenarios they experienced.

    \item Most of the participants who experienced a privacy-invasive scenario that made them feel uncomfortable said that the incident did not escalate to cause harm to them or others (74.3\%).
    
    \item While scenarios that restrict worker autonomy, such as prohibiting them from turning off the camera or microphone, are not among the most commonly experienced scenarios, they are associated with the highest reported discomfort. Participants often violated or would violate the employer's autonomy-restricting rules to protect their privacy. 
    
    \item Participants reported experiencing more discomfort with audio privacy invasions than with video privacy invasions (46.2\% versus 28.8\% of video privacy-invasive scenarios).
   
    \item Smart measures (software/hardware settings or features to prevent privacy-invasive scenarios, e.g. virtual background and noise cancelation features) are not widely adopted compared to manual measures (e.g. manually covering the camera and turning on/off the microphone). When used, more participants reported using settings or features to prevent video scenarios (24.0\%) than audio (13.11\%), data (8.4\%), and autonomy (4.2\%) scenarios. 
\end{enumerate}

Our findings provide useful insights to organizations' policymakers and technology designers to identify areas of improvement to provide healthier work environments. In addition, we provide several open questions that can inspire future work.
\section{Background and Related Work}
\subsection{The Shift to WFH}
The concept of remote work is not new \cite{mcalpine18,hayes21,chiru17}. It has gained popularity in the past 40 years as computers, smartphones, Internet connectivity, and collaboration platforms were becoming ubiquitous \cite{chiru17}. Before 2020, WFH was often provided voluntarily, as a form of benefit such as reduced commuting \cite{hayes21}. However, with the advent of the global COVID-19 pandemic in early 2020, many people who held positions that required them to go to a traditional office space regularly suddenly found themselves required (by their employer or lockdown laws) to stay home, and WFH was no longer a benefit, but a mandate.

WFH mandate created disparate experiences for individuals \cite{ford21}. Ford et al. study on software developers' experiences with WFH found that some developers reported being more productive, while others faced significant challenges \cite{ford21}. WFH can impose challenges on some people, such as affording suitable space at home, and the overlap between work-life and home-life, which in turn may affect worker performance and well-being \cite{fukumura21}. There are also legal considerations about WFH practices, privacy, and how the two might impact an individual's livelihood. For example, in the US, the default employment standard is \quotes{employment-at-will,} which means that individuals are not guaranteed work and their employers may choose to terminate employment at any time \cite{varner22}. As a result, workers may feel the need to self-censor their home environments to help align with what they believe are the social expectations of others in their organization to preserve their reputation and employment status \cite{varner22}. Thus, privacy-protecting behaviors may also be employment-protecting behaviors.

After COVID-19, many individuals and organizations found that the benefits of WFH outweigh the drawbacks. Many organizations have changed their policies to allow WFH on an ongoing basis \cite{smite23,christian23,lufkin22}. Thus, understanding the privacy concerns workers face is a key part of creating healthy work environments. Our work is a step in this direction.

\subsection{Technologies that Enabled WFH}
Technologies that allow workers to work remotely, such as video conferencing, unified communication, and productivity monitoring tools, existed well before the onset of the pandemic and were used regularly by those who had WFH arrangements with their employer \cite{rysavy20}. However, the government-issued mandatory COVID-19 lockdown created a need for employers to quickly switch to a new way of working for their entire organization \cite{hodder20}. Video conferencing technologies, such as Microsoft Teams \cite{teams25}, Google Meet \cite{meet25}, and Zoom \cite{zoom25}, became part of workers' daily lives, and mobile phones (often the worker’s personal property) became an ad hoc office phone. Organizations that had unified communication systems leveraged them in new ways to help workers stay connected \cite{rysavy20}.  

As COVID-19 lockdown extended from weeks to months of remote work, employers became increasingly interested in ensuring that their workers were productive, which resulted in a sharp increase in the demand for remote monitoring technologies \cite{mosendz20}. Remote monitoring tools can provide employers with a variety of surveillance functions. These tools may include features ranging from basic time clock functionality (through which the worker can check in and out of their work shift) to remote monitoring tools which are installed on computers to track the time spent on certain applications, documents opened, email read, and keystrokes \cite{nguyen20,vatcha20}. Some remote monitoring tools can be configured to keep a log of all websites a worker visits and take a screenshot of the worker’s computer screen at regular intervals \cite{allyn20,vatcha20}. Furthermore, in addition to monitoring worker productivity based on their interaction with their computer, it is not uncommon for employers to opt to use video conferencing tools as a worker surveillance tool, requiring workers to keep their cameras and microphones on throughout their workday \cite{mosendz20}. Vitak et al. showed that WFH workers have concerns about worker monitoring practices when data collection exceeds the norms and when monitoring limits workers' autonomy \cite{vitak23}.

Our work complements existing work in this realm and adds new insights to what is known about workers' autonomy-invading scenarios and workers' attitudes towards them.

\subsection{WFH and Security \& Privacy Issues}
WFH \textit{security} and \textit{privacy} issues are often interrelated (a security attack can cause a privacy breach). Sometimes both types of issues are combined as \textit{cybersecurity} issues. However, for the sake of clarity, in what follows, we dissect previous work into two categories: security-focused and privacy-focused studies. Our work is considered a privacy-focused study from the human aspect.

\subsubsection{Security-Focused Studies}
Several studies investigated the security attacks that emerged or increased during the COVID-19 lockdown when WFH was mandated. Pranggono and Arabo provided a summary of notable cybersecurity attacks that increased during the COVID-19 period \cite{pranggono21}. Kotak et al. evaluated the various technologies that allow remote work, and introduced a taxonomy of the security threats that these technologies and WFH introduced to organizations \cite{kotak23}. Lallie et al. highlighted key cybercrime incidents with respect to the timeline of key global events during the COVID-19 pandemic \cite{lallie21}. Other studies provided empirical analysis on particular types of attacks, such as Zoombombing \cite{ling21}, phishing \cite{akdemir21,hoheisel23}, and ransomware \cite{lang23}, to list a few. 

Another line of research on security issues and WFH was focused on workers' and organizations' perspectives. Georgiadou et al. investigated workers' security readiness for WFH during the COVID-19 lockdown \cite{georgiadou22}. Their findings showed a lack of security readiness, especially in terms of the human aspect. For example, 53\% of the workers had not received security guidelines regarding WFH from their organizations, which looks even worse knowing that 44\% of them had never experienced WFH \cite{georgiadou22}. Alromaih et al. conducted a qualitative study on 20 WFH workers investigating the phenomenon \quotes{shadow security behavior,} where workers apply security practices that do not comply with the organization's security policy \cite{alromaih24}. They introduced a model for \quotes{personal security} and identified how it interacts with an organization's security model in the WFH setting \cite{alromaih24}. Bispham et al. interviewed six cybersecurity experts on the cybersecurity issues organizations face in the WFH setting during COVID-19 \cite{bispham22}. The research found that while WFH raised new cybersecurity challenges, experts indicated that cybersecurity issues did not undermine effective WFH \cite{bispham22}. The researchers suggested the need for surveys to gain an empirical understanding of cybersecurity issues during and after COVID-19 WFH setting \cite{bispham22}, which is the contribution of our work.

\subsubsection{Privacy-Focused Studies}
A few recent studies focused on privacy issues in WFH setting. Emami-Naeini et al. conducted a survey on privacy concerns and attitudes toward remote communications in work, socializing, and learning contexts \cite{emami21}. They found that privacy issues in remote communications are among the issues people care about and affect their comfort in using conferencing tools \cite{emami21}. They also found that most of the participants felt that they lacked autonomy to choose the conference tool and/or to use the microphone and camera in their remote communications \cite{emami21}. Prange et al. investigated privacy breaches that occurred as a result of video meetings \cite{prange22}. Of the numerous examples of audio and video-based privacy breaches reported by their participants, more than half of them occurred in business-related scenarios, and many related to video conferencing tools broadcasting accidental audio or video sharing of private situations \cite{prange22}. Cheetham et al. interviewed 18 participants to identify how the COVID-19 pandemic impacted people's privacy \cite{cheetham24}. Their results provided high-level themes and showed that workers face accidental or inappropriate disclosure, feel they have limited control over their privacy, and are reluctant to discuss privacy with their employers due to concerns about privacy stigma \cite{cheetham24}. Whitty et al. conducted a study on the human factor in the WFH environment that includes workers' cybersecurity training and the incidents they may face \cite{whitty24}. While the study is not dedicated to privacy and covered both security and privacy issues, it shed light on a few privacy-invasive incidents such as the unexpected appearance of children at work-related meetings \cite{whitty24}. Herder and Gullit examined workers' privacy-protective behavior with respect to \quotes{power distance,} a term used to describe the extent to which less powerful members of an organization accept variations in power distribution \cite{herder22}. They found that the presence of a superior did not affect participants’ reported privacy-protective measures in group of peers meetings as much as group size and familiarity with group members did \cite{herder22}.  

Several privacy-focused studies focused primarily on remote education context. Li et al. and Castelli et al. investigated students' unwillingness to share their cameras in remote classes \cite{li22,castelli21}. Both studies identified privacy as one of the reasons that led students not to share their cameras in remote classes \cite{li22,castelli21}. Students indicated concerns about their appearance, the appearance of their physical space, or the people around them \cite{li22,castelli21}. Cohney et al. surveyed university instructors and administrators, and analyzed the security of a set of the most popular virtual learning platforms \cite{cohney21}. While the study is not solely privacy-focused, the instructor survey shed light on some privacy issues. It showed that nearly half of the instructors have security and privacy concerns, care about recording controls, and some were dissatisfied with the choice of the platform and its configuration by their university \cite{cohney21}.

Technologies designed to protect workers' privacy can fail under certain assumptions. Hilgefort et al. showed that software features that can be used to enhance privacy while using video conferencing tools, in particular, virtual backgrounds (which can be used to obfuscate surroundings) are vulnerable to attack \cite{hilgefort21}. They found that artificial intelligence can be used to effectively remove the blurred virtual backgrounds native to many video conferencing tools, revealing the participants' true physical setting \cite{hilgefort21}. 

While themes related to some types of privacy-invasive scenarios were identified in prior work, our scenario-based survey combined a list of common privacy-invasive scenarios from different categories (audio, video, data, and autonomy), providing a more comprehensive approach that allowed us to identify nuanced and novel insights that none of the previous work identified. For example, previous work identified autonomy-invading scenarios; however, our results provide insight on their prevalence and the reported discomfort associated with them compared to other scenario categories. While previous work surfaced workers' privacy concerns when working from home, our results evaluated to what extent the discomfort resulting from experiencing privacy-invasive scenarios escalates to cause harm to the workers or others, and what types of harm they cause. Such nuanced insights were missing in previous work, although they are essential for informed decisions about WFH policies and arrangements by employers and policymakers to provide healthier work environments. Unlike the samples surveyed in previous work, which included workers and non-workers or focused on the education context (students and/or educators), our sample includes only WFH workers, providing us with deeper insights into the issues workers face while working from home.
\section{Method} \label{sec:method}
We conducted a 214-participant online survey. We obtained ethical approval for the study from the CMU Institutional Review Board (IRB). All participants were presented with an online consent form at the beginning of the survey. 

\subsection{Design}
Our survey consisted of two parts: a screening survey and a main survey. We used the screening survey to select participants who were currently professional workers (at the time of the study) and working from home regularly (one or more days per week, either in full or hybrid mode), in addition to other screening criteria (e.g. working from home for at least three months), which we detailed in the demographics section of the results.

Only those who met our screening criteria were invited to take the main survey and were immediately redirected to the main survey if they agreed. The screening survey is provided in \autoref{app:surveys}.

To design the main survey questions, three members of the research team developed a set of 14 privacy-invasive scenarios that can occur in a WFH setting, listed in~\autoref{tab:scenarios}. Some of these scenarios were inspired by the results of an IRB-approved unpublished 50-participant exploratory survey on WFH privacy behaviors and concerns conducted by some of this paper's authors and others in 2023. The 14 scenarios were divided into four categories: audio, video, data, and autonomy. The audio and video categories (five scenarios for each category, totaling 10 scenarios) presented leaked audio/video scenarios through the subject's microphone/camera from five different actors: the subject, another adult, child, pet, and object (e.g. appliances, books, furniture, artwork). The data category (two scenarios) presented leaked data scenarios through the subject's shared screen. Finally, the autonomy category (two scenarios) presented restricted subject's autonomy to turn off the microphone/camera. All scenarios were presented in the context of a work-related remote call or meeting from home. 

\begin{table*}[t!]
\centering
\footnotesize
\renewcommand{\arraystretch}{1.25}
\caption{The 14 scenarios presented to participants were divided into categories. Scenarios that share a large amount of text are listed once in the shaded rows for each scenario category, starting with three dots (...) that represent the variable part of the scenario that changes according to the individual scenario listed below.}
\label{tab:scenarios}
\begin{tabular}{l|l|p{11cm}}
\toprule
 No. & Code & Scenario \\
\midrule
\rowcolor{gray!30}
\multicolumn{2}{l}{Audio Scenarios} &  {...} was inadvertently picked up by your device’s microphone, and was heard by one or more people on your work-related remote call or meeting.  \\
\midrule
  1 & voice\_you & \textbf{Your voice}  \\ 
\hline 
  2 & voice\_adult & \textbf{The voice of an adult person (other than you) in your home} \\
\hline
 3 & voice\_child &  \textbf{The voice of a child in your home} \\ 
\hline
 4 & sound\_object & \textbf{The sound of an object (e.g. vacuum cleaner, doorbell, etc.) in your home}\\
\hline
 5 & sound\_pet & \textbf{The sound of a pet in your home}\\
\midrule 
\rowcolor{gray!30}
\multicolumn{2}{l}{Video Scenarios} & ... was inadvertently captured by your device’s camera, and was seen by one or more people on your work-related remote call or meeting.   \\
\midrule
 6 & video\_you & \textbf{Video footage of you}  \\
\hline
 7 & video\_adult & \textbf{Video footage of an adult person (other than you) in your home}  \\
\hline
 8 & video\_child & \textbf{Video footage of a child in your home}\\
\hline
 9 & video\_object & \textbf{Video footage of an object (e.g. books, furniture, artwork, etc.) in your home}\\
\hline
 10 & video\_pet & \textbf{Video footage of a pet in your home}\\
\midrule
\rowcolor{gray!30}
\multicolumn{2}{l}{Data Scenarios}  & ... was inadvertently displayed on your device's shared screen, and was seen by one or more people on your work-related remote call or meeting.  \\
\midrule
 11 & computer\_data &  \textbf{Your data (e.g. emails, files, images, file names, etc.)}  \\
\hline
 12 & browsing\_data & \textbf{Your personalized web browsing data (e.g. personalized ads, auto-completed forms/URLs, browser opened tabs, etc.)} \\
\midrule
\rowcolor{gray!30}
\multicolumn{2}{l}{Autonomy Scenarios}  &   \\
\midrule
 13 & cant\_stop\_camera & \textbf{You} wanted to do something urgent privately (e.g. take a medication) at your home, but were not allowed to stop sharing the camera \\
\hline
 14 & cant\_stop\_mic & \textbf{You} wanted to have an urgent private conversation at your home, but were not allowed to stop sharing the microphone \\
\bottomrule
\end{tabular}
\end{table*}

The survey started with an introduction page where we defined WFH as: \begin{quote} \textit{Work From Home (WFH) is a flexible work setting in which employees perform their duties including collaborative work with others, such as meeting, teaching, etc. remotely from their own homes, mostly using information and communication technologies (e.g. Zoom, Microsoft Teams, Skype, etc.), instead of from a dedicated office space provided by employers.}\end{quote} 
Next, we presented a set of questions about the participant's WFH setup such as whether the participant has a dedicated space they use to work from home, and an open-ended question to describe it. We then presented the scenarios listed in~\autoref{tab:scenarios} in a matrix format, divided by categories, where each category was presented on a separate page of the survey. The scenarios were listed in the matrix rows, while the possible answers for each scenario were presented in the matrix columns with the following options: \quotes{Experienced it and felt very uncomfortable,} \quotes{Experienced it and felt somewhat uncomfortable,} \quotes{Experienced it and felt comfortable,} and \quotes{Never experienced it.} We randomized the order of the scenario matrices for each participant to mitigate the learning and fatigue effects. 

To further characterize WFH scenarios that cause discomfort, after all scenario questions were answered, we presented participants with a list of all the scenarios that they indicated made them feel uncomfortable. Then, we asked participants to select the most memorable scenario and then asked them follow-up questions about the one selected scenario. The follow-up questions included an open-ended question to describe the incident in more detail, and several multiple-choice questions including whether the discomfort escalated to cause harm to themselves or others. If the participant did not report any scenario that made them feel uncomfortable, we asked them to describe a scenario that would make them feel uncomfortable if they were to experience it while working from home and why. The survey concluded with demographic questions.

To avoid security and privacy priming, we did not mention the word privacy or security to participants at any point before or during the survey. The survey was conducted using Qualtrics \cite{qualtrics24}, an online survey platform. The complete screening and main survey questions are included in \autoref{app:surveys}. 

The final survey was refined after conducting two pilot surveys with a total of 19 completed surveys. Pilot surveys are not included in our results.

\subsection{Recruitment}
Our sample size choice (214 participants) was motivated by balancing cost-effectiveness and acceptable statistical power for scientific research \cite{serdar21}. A post hoc power analysis using the G*Power statistical tool \cite{gpower24} for the Chi-square ($\chi^2$) test family with Effect size 0.3 (medium); $\alpha$ err prob = 0.05; total sample size: 214; Df = 1, gives us a Power (1 - $\beta$ err prob) = 0.99, suggesting that 214 provides sufficient power for our Chi-square statistical tests.

We recruited gender-balanced survey participants using Prolific \cite{prolific24}, a research-oriented online participant recruitment platform. We required participants to be at least 18 years old, reside in the United States, able to read and write in English, and have at least a 90\% approval rate in Prolific. We paid participants \$1.00 for completing the screening survey and \$3.75 for completing the main survey. The survey was published on November 14, 2023, and closed in the same day after we reached 228 responses.

\subsection{Analysis} \label{sec:analysis}
We used mixed methods to analyze our data. We analyzed quantitative data using descriptive statistics. To test significance, we used the Chi-square test ($\chi^2$) \cite{mchugh13}, given that no more than 20\% of the expected frequencies are less than 5 and none of the cells are less than 1 \cite{kim17}. We used an alpha level of .05 for our statistical tests. All tests were computed using the R statistical tool \cite{rproject24}. We computed the Chi-square effect size using Cramer's V ($\alpha_c$).

To qualitatively analyze open-ended responses, two researchers used Template Analysis, a style of thematic analysis that combines both inductive and deductive coding, with an emphasis on hierarchical coding without specific prescription regarding the number of levels required and what the levels represent \cite{king24,brooks14}. The first researcher acted as the main coder and the second as a reviewer. Both researchers are trained and have experience conducting qualitative analysis. Moreover, a single coder is deemed acceptable in the Template Analysis method and HCI research \cite{king24,mcdonald19}. Both researchers first familiarized themselves with the data by reading the responses. Next, the first researcher created an initial codebook (the template), and then coded the whole dataset. The first researcher updated the codebook during the coding process. After the first researcher finished coding the entire data set, the second researcher reviewed the coding applied by the first researcher. Both researchers discussed and resolved disagreements and adjusted the codebook and the coding throughout the process as needed.

For analysis purposes and throughout the paper, a scenario is considered uncomfortable if the participant reported that they felt either \quotes{very uncomfortable} or \quotes{somewhat uncomfortable} when they experienced it. We grouped these categories for analysis for reasons deemed appropriate to collapse categories \cite{vaus14,dusen20}: to make them more relevant to our research problem and to highlight patterns, especially as we combined relevant categories with low frequencies. The \quotes{very uncomfortable} category has low frequencies (see \autoref{tab:prevalence} in \autoref{app:more_results} for details). The scope of our exploratory study does not aim to investigate the severity of discomfort, thus treating the two categories (i.e. \quotes{very uncomfortable} and \quotes{somewhat uncomfortable}) separately will not add new information to our findings.
\section{Results}
We first describe our participants. We then describe the prevalence of privacy-invasive scenarios reported by participants, the measures participants took to prevent the presented privacy-invasive scenarios from happening (or happening again), and the reasons for not taking measures if they did not take any. Finally, we highlight further characteristics of the scenarios that caused discomfort. We refer to the participants as P followed by their ID (P\_\#). Finally, we provide frequencies of the main qualitative themes to offer a sense of their prevalence in our dataset.

\subsection{Participants} \label{sec:demographics}
We screened 517 participants. Out of those, 244 met the inclusion criteria and were invited to take the main survey, and 228 completed the main survey. We excluded 14 invalid responses: 5 showed a pattern of very low-quality answers in the open-ended questions indicating inattention; 2 reported they do not share camera, microphone, or screen in any of their meetings in response to screening Q.15--Q.17; 3 duplicate responses; 2 finished the main survey in less than 3 minutes (well below the average time to complete the survey); one response had a technical error; and one reported in the comment at the end of the survey that they misunderstood some questions (our survey purposefully did not have a back button to edit the answers). We ended up with 214 completed and valid responses included in our analysis.

Of the 214 participants in the final dataset, 116 (54.2\%) were males, 94 (43.9\%) were females, 3 (1.4\%) were non-binary, and one preferred not to answer. Their ages ranged from 18 to 74 years, and the majority (69.2\%) fell between 25 and 44 years old. In compliance with our screening requirements, all participants were professional workers (full-time or part-time employees, or self-employed, freelancers, or business owners); had been working from home one or more days per week for at least 3 months (at the time of the survey); were not performing their reported jobs through crowdworking platforms such as Prolific and MTurk\footnote{These differ from the traditional work model and relationships (e.g. with managers and colleagues) that we assumed when we developed our survey's scenarios.}; were communicating with others via conferencing tools such as Zoom at least once every few WFH weeks; were using a device that has a screen (smartphone, tablet, laptop or PC) and were using video conference software (e.g. Zoom) to facilitate WFH. 

Our participants were employed in a wide range of sectors. However, there were a few dominating sectors:  \quotes{Information and communication Technology} (22.4\%), \quotes{Financial} (13.6\%), \quotes{Sales and retail} (12.1\%), and \quotes{Health} (10.7\%). More than a third of the participants (31.3\%) reported \quotes{Other.} Our participants' job titles, provided in an open-text entry, covered a wide range of jobs and levels of seniority. For example, the job titles associated with the most reported sector \quotes{Information and communication Technology} include: \quotes{Administrative Assistant,} \quotes{customer service,} \quotes{IT Tech,} \quotes{Analyst,} \quotes{Senior Data Analyst,} \quotes{Developer,} \quotes{Project Manager,} and \quotes{Consultant.} See \autoref{tab:wfh_context}, \autoref{tab:conf_tools_usage},\autoref{tab:demographics}, and \autoref{tab:wfh_setting} in \autoref{app:more_results} for more demographic and WFH context details.

Most of our participants reported having been working from home for a fairly long time, with 79.4\% reported having been working from home at least one day per week for at least two years. Regarding the frequencies of sharing (turning on) devices (camera, microphone, and screen), our participants reported that, for all or most of their meetings, 149 (69.6\%) shared microphone, 102 (47.7\%) shared camera, and 28 (13.1\%) shared screen. Most participants were living with other people (e.g. spouse, children, parents, roommates, etc.), while only 31 (14.5\%) reported that they were living with \quotes{no one.} 170 (79.4\%) participants said they have a dedicated \quotes{space} to work from home; however, when we asked them to describe the space in an open-ended question, only 59 (27.6\%) mentioned having a dedicated \quotes{office room,} and the rest were working from places other than an office room, such as bedroom (51), secondary bedroom (34), living room/living room-table (35), dining room/dining-table (14), kitchen/kitchen-table (10). See \autoref{tab:wfh_setting} in \autoref{app:more_results} for further details about participants' WFH setting.

\subsection{Prevalence of Privacy-Invasive Scenarios} \label{sec:prevalence}
Our results show that experiencing privacy-invasive scenarios while working from home is common, where 201 (93.9\%) participants experienced at least one of the 14 presented scenarios, and 140 (65.4\%) felt uncomfortable during their experience for at least one of the scenarios. The average number of scenarios the participants experienced is 5.2, and the average number of scenarios that made the participants feel uncomfortable is 2.2 scenarios. See \autoref{tab:prevalence} and Figures \autoref{fig:prevalence}, \autoref{fig:freq_exp}, and \autoref{fig:freq_uncomf} in \autoref{app:more_results} for further details and a breakdown of the results for the scenarios our participants experienced and how they felt about them.

\par
\begin{figure}[t!]
\centering
\includegraphics[width=0.6\textwidth]{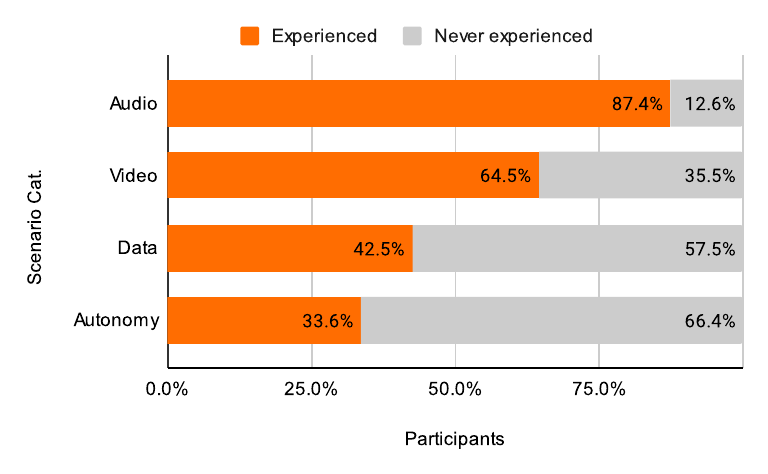}
\caption{Participants who experienced one or more scenarios in a scenario category as a percentage of \textit{all participants}.} 
\label{fig:scenario_prevalence_grouped}
\end{figure}

 We first look at the scenarios experienced by the largest number of participants (regardless of whether they caused discomfort or not), aggregated by category. As~\autoref{fig:scenario_prevalence_grouped} illustrates, audio scenarios are the most commonly experienced scenarios, with 187/214 (87.4\%) participants experiencing at least one audio scenario. The most experienced types of audio scenarios (from the most experienced to the least experienced) were related to leaked sound or voice of an object, a pet, the subject, an adult, and a child. On the other hand, autonomy scenarios are the least experienced scenarios, with 72/214 (33.6\%) participants having experienced at least one autonomy scenario. A breakdown of results for each scenario is presented in \autoref{fig:scenario_prevalence_dissected} in \autoref{app:more_results}. 

\par
\begin{figure}[t!]
\centering
\includegraphics[width=0.6\textwidth]{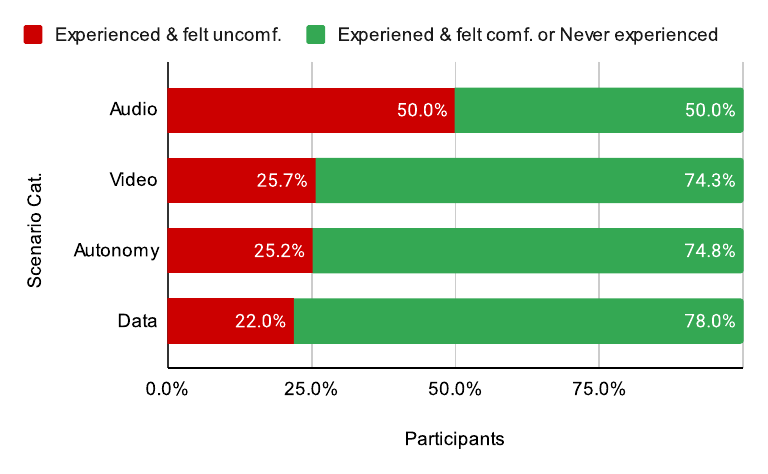}
\caption{Participants who experienced one or more scenarios in a scenario category and felt uncomfortable as a percentage of \textit{all participants}.} 
\label{fig:discomfort_prevalence_grouped}
\end{figure}

 Second, we focus on the scenarios that caused the most discomfort. As shown in~\autoref{fig:discomfort_prevalence_grouped}, audio scenarios were most often reported as causing discomfort, with discomfort reported by 107/214 (50\%) participants for one or more audio scenarios. A breakdown of results for each scenario is presented in \autoref{fig:discomfort_prevalence_dissected} in \autoref{app:more_results}. 

\begin{figure}[t!]
\centering
\includegraphics[width=0.6\textwidth]{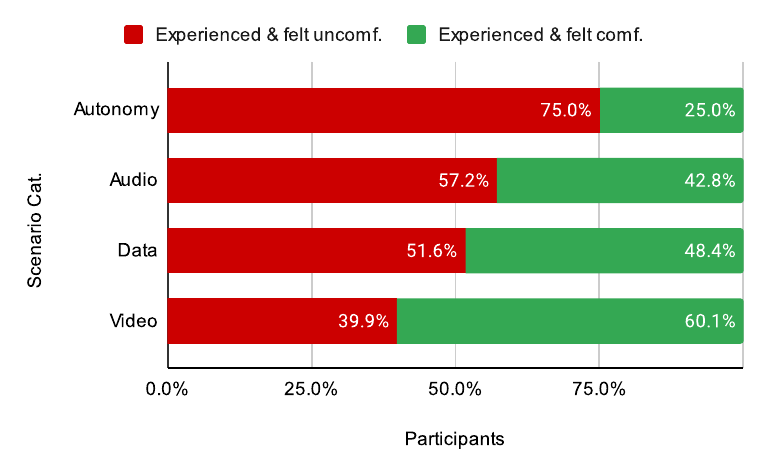}
\caption{Participants who experienced one or more scenarios in a category and felt uncomfortable as a percentage of the \textit{participants who experienced one or more scenarios of that category}.}
\label{fig:ratio_discomfort_prevalence_grouped}
\end{figure}
\par Finally, we present results showing which aggregated scenario categories caused the most discomfort to those who experienced them. That is, the percentage of those who experienced discomfort to the total number of people who experienced one or more scenarios from that category. As~\autoref{fig:ratio_discomfort_prevalence_grouped} illustrates, the autonomy scenario category has the highest percentage, with 54/72 (75.0\%) of those who experienced at least one autonomy scenario reporting that they felt uncomfortable, suggesting that autonomy scenarios are the most privacy-invasive scenarios when they occur. Audio scenarios come next with 107/187 (57.2\%), while video scenarios remain the least reported scenario category that caused discomfort with 55/138 (39.9\%) participants. A breakdown of the results for each scenario is presented in \autoref{fig:ratio_discomfort_prevalence_dissected} in \autoref{app:more_results}. 

\autoref{tab:cat_agg_summary} in \autoref{app:more_results} provides a breakdown of the numbers summarized in~\autoref{fig:scenario_prevalence_grouped},~\autoref{fig:discomfort_prevalence_grouped}, and~\autoref{fig:ratio_discomfort_prevalence_grouped}.

\subsection{Measures Adoption} \label{sec:measures}
At the end of each scenario matrix question (scenario category), we asked the participants if they took any protective measures to prevent the presented scenarios from happening (if they never experienced any) or happening again (if they experienced any). We asked the participants who reported that they took measures to describe the protective measures they took \quotes{including any new software you started using, changes to software settings, changes to your physical workspace, or anything else.} In what follows, we summarize the adoption rate of privacy protection measures and the types of measures our participants took.

\subsubsection{Measures' Adoption Rate} \label{sec:measures_rate}
As shown in~\autoref{tab:measures_by_exp}, our results show that participants who experienced scenarios in a category were more likely to take protective measures. To test the significance of the measures' adoption rate between those who experienced one or more scenarios of a particular category or never experienced any, we computed the Chi-square test in each scenario category. The difference in measures' adoption rate between those who experienced at least one scenario of a category and those who never did is statistically significant in all categories except the audio category (audio: $\chi^2$ = 0.989; p-value = 0.3198; $\alpha_v$ = 0.0537), (video: $\chi^2$ = 5.1425; p-value = 0.023; $\alpha_v$ = 0.145), (data: $\chi^2$ = 7.142; p-value = 0.007; $\alpha_v$ = 0.173), (autonomy: $\chi^2$ = 19.866; p-value $<$ 0.001; $\alpha_v$ = 0.292).

Furthermore, whether the participants experienced any of the scenarios in a particular category or not, in all scenarios except autonomy scenarios, nearly half of the participants said that they adopted measures to prevent audio (57.0\%), video (48.6\%), and data (44.4\%) scenarios; while only 22.4\% of participants reported that they adopted measures to prevent autonomy scenarios. 

\begin{table*}[t!]
\centering
\footnotesize
\caption{Number of participants who took and did not take protective measures. We show all participants in the left section and then break this down into those who experienced scenarios in that category (middle) and those who never did (right). As the table shows, those who experienced scenarios in a category were more likely to take protective measures.}

\resizebox{\textwidth}{!}{ 
\begin{tabular}{p{1.5cm}|ll|ll||ll|ll||ll|ll}
\toprule 
\multirow{2}{*}{\makecell{Scenario\\Cat.}} & \multicolumn{4}{c}{\makecell[c]{All participants\\(Experienced or Never experienced)}} 
                               & \multicolumn{4}{c}{Experienced}    
                               & \multicolumn{4}{c}{Never experienced} 
                               \\
\cline{2-13}
                               & \multicolumn{2}{c}{Took measures} & \multicolumn{2}{c}{Didn't take} 
                               & \multicolumn{2}{c}{Took measures} & \multicolumn{2}{c}{\makecell[l]{Didn't take}} 
                               & \multicolumn{2}{c}{Took measures} & \multicolumn{2}{c}{\makecell[l]{Didn't take}}
                                \\
\hline 
Audio       & 122/214 & (57.0\%) & 92/214 & (43.0\%) 
            &  109/187 & (58.3\%) & 78/187  & (41.7\%) 
            &  13/27 & (48.1\%) & 14/27 & (51.9\%) 
             \\ 
\hline 
Video       & 104/214 & (48.6\%) & 110/214 & (51.4\%) 
            & 75/138 & (54.3\%) & 63/138 &  (45.7\%)  
            & 29/76 & (38.2\%) & 47/76 & (61.8\%)  
            \\
\hline 
Data        & 95/214 & (44.4\%) & 119/214 & (55.6\%) 
            & 50/91 & (54.9\%) & 41/91 & (45.1\%)  
            & 45/123 & (36.6\%) & 78/123 & (63.4\%) 
             \\
\hline 
Autonomy    & 48/214 & (22.4\%) & 166/214 & (77.6\%) 
            & 29/72 & (40.3\%) & 43/72 & (59.7\%) 
            & 19/142 & (13.4\%) & 123/142 & (86.6\%)  
             \\ 
\bottomrule
\end{tabular}
}
\label{tab:measures_by_exp}
\end{table*}
\subsubsection{Types of Measures Taken} \label{sec:measures_type}
We identified four overarching themes that summarize the types of protective measures our participants took to prevent the privacy-invasive scenarios: device control, changes to the surrounding space, involving others, and vigilance. In what follows, we elaborate on each theme. Note that the listed measures are non-exclusive.

\paragraph{\textbf{Device control.}} 
Participants mentioned measures centered around controlling the shared device (microphone for voice, camera for video, and screen for data), mostly in relation to audio and video scenarios. 

To prevent audio scenarios, most of the participants control the microphone. The most mentioned microphone control measures were keeping the microphone muted unless they needed to talk (43) and using mute whenever they needed to prevent unwanted audio from being transmitted (12) as described by P\_220: \quotes{I try to keep myself muted unless I'm talking. I'm usually pretty good about muting/unmuting.} Few participants (16) explicitly mentioned using software or hardware settings or features to control the microphone. For example, using noise cancelation features built into conference software (6) or headset (6) \quotes{to only pick up my voice and minimize any outside noise} (P\_043), using \quotes{push-to-talk} to provide temporal unmute by pressing a key or button (3), or configuring the mute setting to be on by default (2). 

In the same vein, to control video scenarios, participants control the camera. The most mentioned camera control was covering the camera (23), e.g. with a tape or slider as P\_068 who places \quotes{a Post-It note over my camera lens so that I would have to physically take it off before appearing on camera,} keeping the camera off unless needed (8) and turning off the camera whenever needed (6). Few participants (25) explicitly mentioned using software settings to control their camera, such as the virtual or blurred background features available in most conference software (22), or configuring the camera off as their default setting (3). 

For data scenarios, unlike audio and video scenarios, controlling the shared screen is less practical. However, 3 participants mentioned that they limit screen sharing unless necessary: \quotes{I try not to share my screen and do what I can to avoid it} (P\_097). Few participants (8) mentioned software settings, features, or tools to control what is shown on the screen. For example, 5 mentioned window sharing instead of full-screen sharing as P\_076 explained: \quotes{ensure that I don't share my entire screen, but only share the specific program or tab that needs to be shared,} 2 mentioned using private browsing, one mentioned using software that allows to \quotes{share a specific section of my screen} (P\_074), one mentioned auto-deleting the history, and one mentioned using an ad blocker. 

Autonomy scenarios are scenarios in which workers are required to continuously share their microphone or camera. Nonetheless, some participants mentioned that they control their microphone or camera as preventative measures for autonomy scenarios, even though this is prohibited by their employer. For example, participants mentioned turning their microphone off (7) or keeping mute unless needed (4). Similarly, the participants mentioned turning off their camera (6), covering their camera (2), or keeping the camera off unless needed (1). For example, P\_027 explained: \quotes{if i have a true emergency and need to speak to someone - i will simply turn my camera and microphone off for a minute.} This indicates a tension and conflicting situation between the employer's stated requirements in the scenarios (disallowing workers from turning off their microphone or camera) and what some participants do in such situations as P\_195 expressed: \quotes{I decided that I am an adult and if I need to leave my desk or mute my microphone it won't be the end of the world, if my job wants me to be productive as an adult then I need to be treated as an adult.}

\paragraph{\textbf{Changes to the surrounding space.}}
Changes to the surrounding space, either physical or virtual space (the latter for data scenarios), were mentioned across all types of scenarios. Changes in physical space included closing the door, mentioned by 27 to prevent audio scenarios, and by 15 to prevent video scenarios, while 2 participants mentioned it in relation to autonomy scenarios to mitigate the reasons they might want to stop the camera or microphone. Using a door sign to signal that they were in a meeting was also mentioned by a few participants in the audio (4), video (1), and autonomy scenarios (2). For example, P\_068 explained that they \quotes{put a `Please don't knock' sign on my door.} To prevent data scenarios, participants mentioned changes to the virtual space that are equivalent to closing the door in the physical space, such as closing or minimizing the browser, apps, or tabs (32). P\_050 explained that they \quotes{always make sure to close everything before meetings.}  

Clearing the physical background was mentioned in relation to audio and video scenarios. For audio scenarios, removing sources of background noise from the physical space was mentioned by 15 participants, such as ensuring that the \quotes{TV is turned down} (P\_138) and \quotes{turn off Google doorbell alerts and other sounds} (P\_224). For video scenarios, 23 participants mentioned tidying or clearing the physical background or workspace, such as \quotes{rearranged my office} (P\_167 ) and \quotes{just a blank wall behind me} (P\_136). For data scenarios, participants mentioned equivalent changes to the virtual space, such as clearing browsing history or cookies or logging out of accounts (9), and opening a new browser tab (3).

The participants also mentioned the use of a designated or quiet physical space to prevent audio (6), video (9), and autonomy scenarios (1). Changing the place altogether was mentioned by 4 participants in the audio scenarios such as P\_213 who would \quotes{leave the house when it's getting clean (and work elsewhere),} and by 4 in autonomy scenarios, but none of the participants mentioned changing the place in video scenarios. Equivalently, using a designated space or changing the place to prevent data scenarios were expressed through virtual means. For example, 24 participants mentioned using a separate device for work (including two participants who mentioned virtual machines) such as P\_091 who explained: \quotes{I do not mix my business devices with my personal devices.} Participants also mentioned using a second monitor (6) to \quotes{move certain tabs to a different monitor that I don't share} (P\_196), a separate browser (5), and a separate account for work (4).

\paragraph{\textbf{Involving others.}} 
Participants mentioned measures that require involving others who live in their homes to prevent audio, video, and autonomy scenarios, but not data scenarios. This included involving adults, children, and pets. For example, in audio scenarios, 25 participants mentioned measures involving others at home, for example, to \quotes{Ask the people to be quiet during my meeting} (P\_011) or \quotes{to watch the cat} (P\_012). Similarly, in video scenarios, 10 participants would inform others mainly \quotes{to not come in during certain hours} (P\_132) or \quotes{to knock or text my phone first} (P\_137). Involving others at home was also mentioned in autonomy scenarios by 8 participants, for example \quotes{to remind them that I should not be disturbed} (P\_091). 

In limited cases, participants mentioned involving people at work-related meetings or calls. This was mentioned by 3 in relation to audio scenarios, for example: \quotes{if there is something going on outside my home such as construction. I usually just apologize and let them know what is going on and when they can expect the noise to stop} (P\_091). 3 participants in autonomy scenarios would inform people at work when they need to leave. 

Involving pets including removing or managing pets such as \quotes{make sure the dogs are comfortable or put away for that time} (P\_165), is a common measure that is repeatedly mentioned mostly in relation to audio (21), and to a lesser extent video (3). 

\paragraph{\textbf{Vigilance.}} 
Vigilance was expressed through being mindful of the physical or virtual surroundings, mostly in relation to video, data, and autonomy scenarios. In video scenarios 23 participants mentioned being mindful of the camera position and what it shows as P\_077 described: \quotes{I make sure that my private living area cannot be seen.} Only 2 participants in audio scenarios mentioned being mindful of their surroundings. In autonomy scenarios, mindfulness was expressed through planning ahead of time and time management, as mentioned by 15 participants. For example, the participants repeatedly mentioned comments such as \quotes{I make sure door is closed during meetings and have everything I need} (P\_049), \quotes{I try to get everything done} (P\_043), and \quotes{I try to go to the bathroom} (P\_024), before the meeting time. Equivalently, in data scenarios, participants mentioned being mindful of what is shown on the screen (20), mindful of their activity on the work device (19), or their activity during the work or meeting time (4).

Vigilance was also expressed by checking the device status. In audio scenarios, checking the microphone status was mentioned by 11 participants, with 6 of them mentioning repeated checks at various frequencies: \quotes{double check} (P\_112), \quotes{multiple times} (P\_033), \quotes{triple-check} (P\_166), and \quotes{constantly keep ensuring} (P\_185). In video scenarios, checking the camera status was also mentioned, but at a lower rate (possibly because if the camera is on, this is immediately reflected on the screen, unlike the mic), where only 3 mentioned it, with only one of those participants mentioning repeated camera checks.
\begin{table*}[t!]
\centering
\footnotesize
\caption{Frequencies of the identified themes that form the reasons for not taking protective measures to prevent the privacy-invasive scenarios.}

\begin{tabular}{l|l|l|l|l}
\toprule 
    Reasons' Themes     & Audio & Video  & Data  & Autonomy \\ 
\midrule 
    Never experienced  & 10  & 11 & 31 & 73 \\
    \hline 
    Oblivious to measures & 23 & 31 & 54 & 20 \\
    \hline 
    Understanding work environment & 15 & 12 & 3 & 21 \\
    \hline 
    \makecell[l]{Nothing can be done} & 22 & 7 & 2 & 18 \\
    \hline 
    Unbothered      & 42 & 51 & 24 & 29 \\
    \hline 
    Nothing to hide & 2 & 19 & 26 & 0 \\
\bottomrule
\end{tabular}
\label{tab:reasons_no_measures}
\end{table*}
\subsection{Not Adopting Measures} \label{sec:no_measures}
At the end of each scenario matrix question (scenario category), we asked participants who reported that they did not take any measures to prevent the presented scenarios from happening (if they never experienced any) or happening again (if they experienced any) about the reasons for not taking any measures. 

As shown earlier in~\autoref{tab:measures_by_exp}, our results show that many participants, whether they experienced any of the scenarios in a particular category or never did, said they did not take any measures to prevent the scenarios from happening (or happening again): audio (43.0\%), video (51.4\%), data (55.6\%), and autonomy (77.6\%) scenarios. 

Participants who said they did not take any protective measures provided a variety of reasons for not doing so. We identified six prominent themes that encompass most of these reasons: never experienced, oblivious to measures, understanding work environment, nothing can be done, unbothered, and have nothing to hide. In what follows, we elaborate on each theme. 

Since these themes are universal across all scenario categories, for readability, we refer the reader to~\autoref{tab:reasons_no_measures} for themes' frequencies. Note that the reasons for not taking measures are non-exclusive.

\paragraph{\textbf{Never experienced.}}
One intuitive reason for not taking any measure for some or all scenario categories is that the participant never experienced the scenarios. This was mostly mentioned in relation to autonomy scenarios, which are the least experienced scenarios as P\_103 explained: \quotes{I have not had this occur so it is not a concern for me,} and to a lesser extent in other scenarios: data, video, and audio, respectively.

\paragraph{\textbf{Oblivious to measures.}}
Many participants who said they did not take any measures appeared (based on their answers) to be taking preventative steps, but did not realize that those steps were a type of preventative measure we were asking about. For example, P\_040 said they did not take measures to prevent data scenarios, however, when asked about the reason, she elaborated: \quotes{I only use my work computer for work-related emails and web browsing so I do not feel the need for protective measures.} Similarly, in audio scenarios, P\_131 said that she \quotes{found a quiet space/designated area, so no need for protective measures.} The steps taken by these participants, using a separate computer for work purposes or using a quiet or designated space, are indeed measures to prevent privacy-invasive scenarios. 

\paragraph{\textbf{Understanding work environment.}}
An understanding and friendly work environment was mentioned across all scenarios as a reason to not adopt measures, notably in relation to autonomy scenarios, where this theme also includes flexible or no work policies that restrict worker's autonomy as summarized by P\_168: \quotes{My organization has always been understanding that we might need to turn off our microphone or camera. There's been no negative repercussions around doing so,} and to a lesser extent in other scenarios: audio, video, and data.

\paragraph{\textbf{Nothing can be done.}}
Many participants across all scenarios felt there was nothing they could do to prevent the scenarios or that the scenarios were inevitable. Notably, in relation to audio and autonomy scenarios as P\_075 explained why she did not take measures to prevent audio scenarios: \quotes{some things are unpreventable.} 

\paragraph{\textbf{Unbothered.}}
Some participants mentioned that the scenarios did not, or would not, bother them, or expressed they felt that there was no need to take any measures. Some participants did not explain and just noted: \quotes{Not necessary} (P\_157), while some of them followed that with oblivious measures such as \quotes{Never needed to since I use a dedicated room} (P\_181). 

\paragraph{\textbf{Have nothing to hide.}}
Many expressed they have nothing to hide mostly in relation to data and video scenarios, which included nothing sensitive, confidential, offensive, embarrassing, inappropriate, bad, important, or personal as P\_017 elaborated: \quotes{Nothing that has ever been shown was non-work appropriate. Haven't felt the need to adopt protective measures yet.}

\subsection{Characterizing Privacy-Invasive Scenarios} \label{sec:character}
To better understand the factors that make a privacy-invasive scenario uncomfortable, we asked participants to select one of the scenarios (the most memorable) that made them feel uncomfortable and asked them detailed follow-up questions about it. Through these detailed questions, we identified some notable patterns of scenarios that caused discomfort as we will elaborate next (see \autoref{tab:detailed} in \autoref{app:more_results} for further details). For participants who did not report any scenario that made them feel uncomfortable (they either felt comfortable or had never experienced any of the presented scenarios), we asked them to describe an incident that would make them feel uncomfortable and why.

\subsubsection{Reasons for Discomfort}
We first look at the scenarios that the participants selected as the most memorable scenario that made them feel uncomfortable. Of the 140 participants who reported one or more scenarios that made them feel uncomfortable, we find that the most selected memorable scenarios tended to be from the audio category (74/140), and the least selected scenarios are from the video category (13/140). See \autoref{tab:memorable_scenarios} in \autoref{app:more_results} for further details.

\begin{table*}[tbp!]
\centering
\footnotesize
\caption{Frequencies of the identified themes that form the reasons for discomfort when experiencing  privacy-invasive scenarios while working from home.}

\begin{tabular}{p{5cm}|p{5cm}|l|l|l|l}
\toprule 
\multirow{2}{*}{Themes} & \multirow{2}{*}{Sub-themes}	& \multicolumn{4}{c}{Scenario Cat.} \\
\cline{3-6}
                       &  & Audio	    & Video	    & Data   & Autonomy   \\
\midrule 
Affecting the flow of the meeting 
                        & Interrupting; Unexpected; Non-work-related  
                        & 27 & 4  & 2  &  2 \\               
\hline
Affecting the worker's professional image                           
                        & Embarrassing; Unprofessional; Fear of peer judgement; Brought unwanted attention; Inappropriate 
                        & 38 & 13 & 7 & 5  \\                       
\hline
Feeling lack of control over the incident
                        & Heard against will; Seen against will; Cannot stop the meeting; Cannot control  
                        & 12 &  5 & 8  & 6 \\
\hline 
Fear of consequences       & n.a.
                        & 1	& 0         & 0      & 2 \\
\bottomrule
\end{tabular}
\label{tab:reasons_discomfort}
\end{table*}

Our qualitative analysis of the open-ended question about why participants felt uncomfortable (Q.21) revealed four overarching themes for the causes of discomfort: affecting the flow of the meeting, affecting the worker's professional image, feeling a lack of control over the incident, and fear of consequences.~\autoref{tab:reasons_discomfort} lists the frequencies of the identified themes that form the reasons that caused discomfort when experiencing privacy-invasive scenarios while working from home. The audio scenarios have the highest frequencies in three of the four themes.

For the 74 participants who did not experience any scenario that made them feel uncomfortable, the described scenarios that would make them feel uncomfortable were centered on the workers themselves and other adults, with the top scenario categories related to\footnote{Since the scenarios here were provided in response to an open-ended question and not on a list as those who said they experienced at least one scenario that made them feel uncomfortable, some participants provided scenarios that contained multiple categories. Thus, scenario categories in this section are non-exclusive and the counts represent the mentioned scenario categories and not the participants.}: their video (24), their voice (11), an adult's video (11), and an adult's voice (10). Most of the scenarios contained exacerbated details that our scenarios did not include. For example, 26 participants provided scenarios related to inappropriateness, mainly related to being inappropriately dressed or undressed, such as \quotes{If I inadvertently shared my camera and was not dressed appropriately for work.} (P\_155) and \quotes{If i was in the nude thinking my camera was off but it actually wasn't} (P\_143), as well as inappropriate behaviors or conversations.

Looking at the reasons that would make these anticipated scenarios uncomfortable, they are the same reasons that were expressed by those who experienced our scenarios, but with different priorities. Here, we observe that reasons related to interrupting the flow of the meeting are barely mentioned (5), and most of the reasons are centered around affecting the worker's professional image (28/74), feeling a lack of control over the incident (17/74), and fear of consequences (6/74).

\subsubsection{When Discomfort Escalates to Harm} \label{sec:characterize:harm}
Most of the participants 104/140 said the scenario did not escalate to cause harm to self or others, while 36/140 said it caused some amount of harm (35 chose \quotes{small amount of harm} and one \quotes{large amount of harm}). Of the 36 participants who reported harm, 24/36 participants reported audio scenarios, 6/36 data, compared to only 4/36 autonomy, and only 2/36 video scenarios. 

Out of the 36 who reported harm, 34 said that the harm was \quotes{Psychological harm,} one participant reported \quotes{Financial harm} in addition to psychological, one reported \quotes{Physical harm,} and one selected \quotes{Other} and specified it was a \quotes{Reputational harm} (which fits under psychological harm but we keep answers as reported). When we asked them to explain how it caused the harm, the most recurring themes were related to: feeling embarrassed mentioned by 11 participants, unprofessional (4), fear of incident recurrence or paranoia (3), feeling ashamed (2), and anxious (2). P\_187 explained the psychological harm: \quotes{It made me paranoid of the situation repeating and constantly having to make sure to an almost extreme extent that everything is in order before I use my camera.} The one financial harm was due to a one-month suspension from work, the physical harm was in relation to the participant's loud dog barking that he could not stop and explained the harm to others as \quotes{sudden loudness ... which hurt their ears} (P\_105), while the reputational harm was because the incident \quotes{caused some within my organization to talk} (P\_208).

The harm has little relation to whether there were manager(s) presented in the meeting or not. Of the 36 participants who reported harm, almost half (17/36) reported that there were managers at the meeting, while the remaining did not.

\subsubsection{The most shared device and recording} 
Participants reported microphone was shared more often in incidents that caused them discomfort (115/140), compared to 82/140 who shared the camera, and 28/140 who shared the screen. In more than half of the cases (88/140), participants said that those incidents were not recorded, suggesting that even non-recorded incidents still cause discomfort to a large extent.

\subsubsection{Recurrence and location}
Our results suggest that whether the incident occurred once or multiple times, and whether the incident occurred in an office room or in other rooms, have little impact on discomfort. That is, nearly half of the participants (71/140) reported that the incident that caused them discomfort occurred once, while a total of 69/140 reported more than once (\quotes{A few times} or \quotes{Repeatedly}). Similarly, 59/140 reported that the incidents occurred in an office room, while a total of 81/140 in a room other than an office. 
\section{Discussion}
This study aims to provide detailed insights of the privacy-invasive scenarios that workers face while working from home. In what follows, we summarize our key findings, provide practical recommendations, and raise open questions that can motivate future work.

\subsection{WFH Discomfort is Unlikely to Cause Harm, Yet Harmed Minority Should Not be Ignored} \label{sec:harm}
WFH privacy-invasive scenarios are prevalent. Almost all participants experienced at least one scenario. More than half of the participants felt uncomfortable in at least one of the presented scenarios. When asked about harm, most of the participants who experienced discomfort did not find the discomfort caused harm to themselves or others (74.3\%). Almost all of those who reported that the discomfort had escalated to harm said that the harm was psychological and minor. While only 25.7\% of participants who experienced discomfort (16.8\% of all participants) said that it escalated to harm, this is still a considerable minority that should not be ignored. 

Our results suggest that the privacy-invasive scenarios that workers experience while working from home often do not cause harm. Nevertheless, organizations that intend to enforce WFH either in full or hybrid mode need to consider that there might be a harmed minority. One possible suggestion to address this is to involve workers in discussions about WFH policy, listen to the needs of those who might be harmed, and consider ways to mediate this.

\subsection{Rethink WFH Policies that Restrict the Worker's Autonomy} \label{sec:autonomy}
Our results show that autonomy scenarios that limit the subject's autonomy, such as prohibiting them from turning off their camera or microphone during a work-related call or meeting, are the least prevalent scenarios. However, when they are experienced, these scenarios have the highest discomfort rate compared to other scenarios. Moreover, as shown through our participants, many workers reported measures that violated or would violate these rules when needed, despite the stated rules in the autonomy scenarios such as turning off the camera or muting the microphone to protect their privacy, which may result in tension and conflicting situations between workers and employers.  

One might argue that workers should have no expectations for privacy on employer's owned devices and that such autonomy-invading policies are set to protect organizations \cite{shrm24}, e.g. to prevent workers from working in public spaces that can expose confidential information. However, our results show that adherence to such policies may not be always possible, and there are no practical measures to prevent emergencies that may require workers to turn off their camera or microphone. Thus, we suggest that employers rethink policies that limit the worker's autonomy such as disallowing them to turn off their cameras or microphones. For example, we suggest employers discuss and identify the exceptional situations where workers can be exempted from these autonomy-restricting policies, which is inline with Emami-Naeini's recommendation \cite{emami21}. This may release the tension that such policies impose on worker as we observed in some participants' comments.

\subsection{Audio Privacy-Invading Scenarios are More Related to Discomfort and Harm than Video Scenarios} \label{sec:audio}
Our results show that audio scenarios are the most experienced scenarios among our participants. We note that the prevalence of audio scenarios might be in part due to the frequency with which participants shared their microphone as opposed to camera or screen sharing. Regardless, audio scenarios have the second highest discomfort ratio, after the much less frequently experienced autonomy scenarios. When comparing the discomfort ratio in all audio vs. all video scenarios, respectively 239/517 (46.2\%) vs. 98/340 (28.8\%), considerably more audio scenarios caused discomfort than video scenarios (see \autoref{tab:prevalence} in \autoref{app:more_results} for details). Additionally, when we asked participants to select the most memorable uncomfortable scenario from the list of all reported uncomfortable scenarios, the four most reported scenarios were audio scenarios. On the other hand, the least four reported memorable scenarios were video scenarios. Furthermore, looking at the reasons that made the memorable scenarios uncomfortable, we find that audio scenarios appeared more frequently in relation to interrupting or distracting the meeting and affecting the worker's professional image than video and other scenarios. Audio scenarios appeared considerably more than video scenarios in scenarios that escalated to cause harm (24/36 audio vs. 2/36 video).

The aforementioned findings show that audio scenarios are more prevalent, and more related to discomfort and harm, suggesting that they are more difficult to prevent than video scenarios. Future work can better understand the reasons that made audio privacy-invasive scenarios more prevalent, and associated to discomfort and harm compared to video scenarios. 

\subsection{\quotes{Smart} Measures are NOT Widely Adopted Compared to Manual Measures} \label{sec:smart}
Our results show that applying \quotes{smart} privacy-protective measures through software/hardware settings or features that would automate the process (e.g. set the microphone to mute by default instead of manually muting the microphone) or using technical means to replace manual efforts (e.g. use virtual or blurred background instead of manually clearing the physical background) are not widely adopted measures among our participants compared to manual measures. Although we equally primed participants to the two types of measures when we asked them to describe the measures they took \quotes{including any new software you started using, changes to software settings, changes to your physical workspace, or anything else,} still, only a minority of participants who took measures in each category reported smart measures: 16/122 (13.1\%) participants in voice scenarios, 25/104 (24.0\%) in video, 8/95 (8.4\%) in data, while only 2/48 (4.2\%) mentioned smart measures to prevent autonomy scenarios. 

One possible explanation for such phenomenon is a lack of awareness of some smart measures from the first place. For example, Li et al.'s study on the reasons that drive students not to share their cameras in classrooms found that many students were concerned about the appearance of their background or the people around them, suggesting that students lack awareness about smart features on conferencing tools such as virtual background \cite{li22}. Another reason is a lack of awareness that some features are applied by default. For example, some conference tools such as Zoom have the \quotes{Zoom background noise removal} set to \quotes{Auto (automatically adjust noise suppression)} by default. There is no indicator that signals the activation status of this setting in Zoom. Thus, some participants may be oblivious to it. 

It remains unclear what are the reasons for the low usage of smart measures in general. Is it a lack of need, awareness (of their existence or default activation status), convenience, usability, discoverability, skills to use them, or lack of trust in the effectiveness of smart measures in conference tools? We leave these questions to future work.

\subsection{Do Smart Measures Visualization/Indicators Matter?} \label{sec:visualization}
While smart measure adoption was low in general, from this study we observe that smart measures to prevent video scenarios such as virtual or blurred backgrounds which are reflected on the worker's and the meeting participants' screens were more adopted than other types of smart measures that are not reflected on the meeting's screen such as noise cancelation features. Our results show that 25/104 (24.0\%) participants in video scenarios mentioned smart measures compared to 16/122 (13.1\%) participants in audio scenarios.  

This raises an open question that we leave to future work: Does visualization/indicators matter in raising awareness and adoption rate of smart measures in conference tools?

\subsection{Research Questions Revisited}
In what follows we summarize key findings for our main research questions. \par 

\begin{itemize}

\item \textbf{How prevalent is experiencing privacy-invasive scenarios among WFH workers?}

Our results show that WFH privacy-invasive scenarios are prevalent among our participants. Almost all participants (93.9\%) reported that they have experienced at least one of the privacy-invasive scenarios we presented to them. 

\item \textbf{To what extent does experiencing these scenarios cause discomfort?} 

More than half of the participants (65.4\%) felt uncomfortable during at least one of the scenarios they experienced.

\item \textbf{Does discomfort escalate to cause harm to workers or others?}

Most of the participants who experienced discomfort in privacy invasive scenarios said that the discomfort did not cause harm to them or others (74.3\%).  

\item \textbf{What type of harm do workers experience?}

Almost all the participants who reported that the discomfort they experienced escalated to cause harm to them or other said the harm was psychological (34/36) and minor (35/36). 

\item \textbf{What types of scenarios are associated with the highest reported discomfort?} 

Our results show that scenarios that restrict worker autonomy, such as prohibiting them from turning off the camera or microphone are associated with the highest reported discomfort when experienced (75\%). Participants often violated or would violate the employer’s autonomy-restricting rules to protect their privacy.

\item \textbf{What protective measures do some workers take?}

Most participants reported manual measures such as manually covering the camera and turning on/off the microphone. Smart measures (software/hardware settings or features to prevent privacy-invasive scenarios, e.g. virtual background and noise cancelation features) are not widely adopted compared to manual measures. It should be noted that when smart measures are used, more participants reported using settings or features to prevent video scenarios (24.0\%) than audio (13.11\%), data (8.4\%), and autonomy (4.2\%) scenarios.

\item \textbf{Why do some workers choose not to take protective measures?}

Participants reported a variety of reasons for not taking protective measures across all scenario categories with varying rates. The reported reasons include: never experiencing the scenario, having an understanding work environment, feeling that there is nothing they can do to prevent the scenario, or not feeling bothered about experiencing it. Interestingly, some participants were oblivious to measures they were already taking that may help prevent privacy-invasive scenarios. That is, they reported not taking measures, but their open-ended answers to the reasons for not doing so revealed that they are taking measures such as using a separate device for work and using a quiet or designated space for working from home.
\end{itemize}

\subsection{Limitations} \label{sec:limitation}
First, we used the Prolific \cite{prolific24} crowdworking platform to recruit participants, and Prolific workers are known to be more technically skilled than the average population. However, Prolific has been shown to provide reasonably generalizable results in security and privacy research \cite{tang22}. Moreover, we screened for Prolific users who were not performing their reported jobs through crowdsourcing platforms such as Prolific and MTurk. This may have skewed our participants towards a certain type of Prolific users. Second, we used self-reported surveys, which are naturally prone to recall bias and social desirability effects. To mitigate recall bias, we asked participants to answer based on their experiences within the last three years. Moreover, in the open-ended questions when we asked participants what protective measures they took, we equally nudged participants to mention manual and smart measures. However, participants may still have trouble accurately recalling incidents or measures within a three-year time frame. Third, we acknowledge a limitation in the follow-up detailed questions on the scenarios that caused discomfort, where we only followed up with the participants about one scenario they reported as the most memorable scenario that made them feel uncomfortable. However, asking participants follow-up questions about multiple scenarios would likely result in survey fatigue that would affect the quality of the data. Fourth, we did not ask the participants what conference software they use. Different software may have different default settings and features. Moreover, some participants may not be fully aware of their software's default settings and features. Fifth, our study does not consider potential nuanced differences between different types of jobs. This can have implications for the type of rules imposed by employers and the consequences of privacy invasions for workers. Finally, our sample is limited to participants from the US. Privacy laws and social norms can vary between jurisdictions, which needs to be considered when interpreting our results outside the US context. Even within the US, privacy regulations and labor laws vary somewhat by state, which may impact workers' experiences.

\section{Conclusion}
In this study, we surveyed 214 workers about their experiences with 14 privacy-invasive scenarios that can occur in a WFH setting. Our findings provide a better understanding of the privacy challenges that WFH workers face and how they address them, which can help identify ways to improve WFH experiences. 
\section{Competing Interests}
No competing interest to declare.

\section{Author Contributions Statement}
E. A. (Conceptualization, Data curation, Formal analysis, Funding acquisition, Investigation, Methodology, Visualization, Writing – original draft, Writing – review \& editing); J. P. (Conceptualization, Data curation,  Methodology, Writing – original draft); M. L. (Conceptualization, Data curation, Formal analysis, Methodology, Writing – original draft); L. F. c. (Conceptualization, Funding acquisition, Methodology, Writing – review \& editing) 

\section{Acknowledgment}
We thank Prof. Marc Dacier from KAUST for feedback on earlier versions of some sections of this paper. We also thank Harish Balaji, Mahith Edula, Rachel Martini, and Zili Zhou for their contributions to the initial survey study during the Usable Privacy and Security course at CMU. Finally, we thank the participants for their time and valuable input. 

\section{Funding}
E. A. acknowledges the financial support of the Ibn Rushd Program at King Abdullah University of Science and Technology (KAUST). This work was supported in part by the Innovators Network Foundation (L. F. C.).

\bibliographystyle{ACM-Reference-Format}
\bibliography{ref}
\appendix
\twocolumn
\small
\section{Appendix A - Surveys} \label{app:surveys}
In this section we include the surveys we used in the study. The subsections and text between [square brackets] were not shown to participants. We include them here for clarity. \autorefappendix{app:recruitment_survey} lists the recruitment materials.~\autorefappendix{app:screening_survey} lists the screening survey, and~\autorefappendix{app:main_survey} lists the main study's survey. 

\subsection{RECRUITMENT MATERIALS} \label{app:recruitment_survey}
\noindent \textbf{Title}: Your Work From Home Experiences

\noindent \textbf{Reward}: \$1.00 (screening) \${3.75} (main) (approximately \${15}/hr)

\noindent \textbf{Estimated completion time}: {15} mins  

\noindent \textbf{Description}: Our research team at Carnegie Mellon University is searching for people to participate in a survey about their Work From Home Experiences. \textbf{Participants should be working from home to participate.}

If you are interested in participating in our survey, please complete this initial screening survey, which should just take about \textbf{4 minutes}. You will be paid \textbf{\$1.00} for completing the screening survey. All responses in the screening survey will be kept confidential.

If you are selected for the study’s main survey, you will be asked if you agree to take the main survey and you will be redirected to the main survey that will last about \textbf{15 minutes}. You will be paid a \textbf{bonus of \$3.75} for completing the main survey through Prolific.

\noindent \textbf{Devices you can use to take this study}: Mobile, Tablet, Desktop


\subsection{SCREENING SURVEY} \label{app:screening_survey}
\subsubsection{CONSENT}~\\
\noindent [consent form is shown here] \newline 
\noindent \textbf{Q.0:} Please enter your Prolific ID. \newline 
Please note that this response should auto-fill with the correct ID. \newline  
\noindent[Open-text entry]

\subsubsection{INTRODUCTION}~\\
Thank you for your interest in our study. Please complete this initial \textbf{4 minute} screening survey. You will be paid \textbf{\$1.00 for completing the screening survey through Prolific.} If you are selected for the study’s main survey, you will be asked if you agree to take the main survey and you will be directed to the main survey that will take about \textbf{15 minutes}. You will be paid \textbf{\$3.75 for completing the main survey through Prolific.} \newline

\noindent \textcolor{red}{\textbf{Note that the survey does not allow you to return back to the previous question once you hit \quotes{Next}. Please make sure you have selected the choice you really want before you click the \quotes{Next} button.}} 

\noindent \textbf{Q.1:} Which of the following best describes your current employment status?
\begin{itemize}
    \item Employee (full-time)
    \item Employee (part-time)
    \item Self-employed / Freelancer / Business-owner
    \item Student
    \item Homemaker
    \item Unemployed, and looking for a job
    \item Unemployed, and not looking for a job
    \item Unable to work
    \item Retired
    \item Other (please specify): [open-text entry]
\end{itemize}

\noindent[Q.2 to Q.8 are displayed if Q.1 answer is (\quotes{Employee (full-time)} OR \quotes{Employee (part-time)} OR \quotes{Self-employed / Freelancer / Business-owner})]

\noindent \textbf{Q.2:} What is your current job title? (e.g. Teacher, Administrative Assistant, Nurse, etc.) \newline 
\noindent[Open-text entry]

\noindent \textbf{Q.3:} In your job that you described in the previous questions, do you perform your work through crowdsourcing platforms such as Prolific, MTurk, etc.?
\begin{itemize}
    \item Yes 
    \item No 
    \item Other (please specify): [open-text entry]
\end{itemize}

\noindent \textbf{Q.4:} What sector do you currently work in?
\begin{itemize}
    \item Pre-university Education 
    \item University Education 
    \item Health 
    \item Information and Communication Technology 
    \item Financial 
    \item Industrial 
    \item Agricultural 
    \item Sales and retail 
    \item Petrochemical 
    \item Other (please specify): [open-text entry]
\end{itemize}

\noindent \textbf{Q.5:} Which of the following best describes the organization you currently work in?
\begin{itemize}
    \item Privately-held organization 
    \item Publicly-traded organization 
    \item Government organization 
    \item Educational organization
    \item Not-for-profit organization 
    \item Other (please specify): [open-text entry]
\end{itemize}

\noindent \textbf{Q.6:} Approximately, how long have you been working in your job described above?
\begin{itemize}
    \item Less than a year 
    \item One year 
    \item 2 years 
    \item 3 years 
    \item 4 years 
    \item 5 years 
    \item More than 5 years 
    \item Other (please specify): [open-text entry]
\end{itemize}

\noindent \textbf{Q.7:} On average, how many days per week do you work overall?
\begin{itemize}
    \item One day per week 
    \item 2 days per week 
    \item 3 days per week 
    \item 4 days per week 
    \item 5 days per week 
    \item 6 days per week 
    \item 7 days per week 
    \item Other (please specify): [open-text entry]
\end{itemize}

\noindent \textcolor{red}{\textbf{Please think only about your current job that you just reported in the previous section, and answer all the following questions in that context.}}

\noindent \textbf{Q.8:} \textbf{Do you currently} work from home regularly (one or more days per week, either in full or hybrid mode), where you perform your duties including collaborative work with others, such as meeting, teaching, etc. remotely \textbf{from your own home}, mostly \textbf{using information and communication technologies (e.g. Zoom, Microsoft Teams, Skype, etc.)}, instead of from a dedicated office space provided by employers?
\begin{itemize}
    \item Yes 
    \item No 
\end{itemize}

\noindent [\textbf{Q.9 - Q.14} are displayed if \textbf{Q.8} answer is \quotes{Yes}]

\noindent \textbf{Q.9:} On average, out of the [Q.7 answer] that you work, how many of those days are you working from home for all or part of the day?
\begin{itemize}
    \item One day per week 
    \item 2 days per week 
    \item 3 days per week 
    \item 4 days per week 
    \item 5 days per week 
    \item 6 days per week 
    \item 7 days per week 
    \item Other (please specify): [open-text entry]
\end{itemize}

\noindent \textbf{Q.10:} Approximately, how long have you been working from home (either in full or hybrid mode)?
\begin{itemize}
    \item Less than 3 months 
    \item From 3 to 11 months 
    \item One year 
    \item 2 years 
    \item 3 years 
    \item 4 years 
    \item 5 years 
    \item More than 5 years 
    \item Other (please specify): [open-text entry]
\end{itemize}

\noindent \textbf{Q.11:} Typically, when working from home, approximately, how often do you need to communicate with others via conferencing tools such as Zoom or Microsoft Teams?
\begin{itemize}
    \item At least once every work from home day 
    \item At least once every few work from home days 
    \item At least once every work from home week 
    \item At least once every few work from home weeks 
    \item At least once every work from home month 
    \item Less than once every work from home month 
    \item Other (please specify): [open-text entry]
\end{itemize}

\noindent \textbf{Q.12:} Typically, when working from home, on a day where you communicate with others, approximately, how much of the day do you spend communicating with others via conferencing tools such as Zoom or Microsoft Teams?
\begin{itemize}
    \item Less than one hour per work from home day 
    \item At least one hour per work from home day 
    \item At least 2 hours per work from home day 
    \item At least 3 hours per work from home day 
    \item At least 4 hours per work from home day 
    \item More than 4 hours per work from home day 
    \item Other (please specify): [open-text entry]
\end{itemize}

\noindent \textbf{Q.13:} Typically, when working from home, which of the following devices do you use? (select all that apply)
\begin{itemize}
    \item Mobile smartphone 
    \item Mobile traditional (non-smart) phone 
    \item Landline phone 
    \item Tablet 
    \item Laptop 
    \item Personal Computer (PC) 
    \item Other (please specify): [open-text entry]
\end{itemize}

\noindent \textbf{Q.14:} Typically, when working from home, which of the following technologies do you use to facilitate remote work from home?  (select all that apply)
\begin{itemize}
    \item Video conferencing (e.g. Zoom, Google Meet, Microsoft Teams, Skype, etc.) 
    \item Phone (audio only) 
    \item Email messaging 
    \item Instant messaging applications (Slack, WhatsApp, etc.) 
    \item Mobile Short Messaging Service (SMS) messaging 
    \item Other (please specify): [open-text entry]
\end{itemize}

\noindent[\textbf{Q.15 - Q.17} are displayed if \textbf{Q.14} answer is \quotes{Video conferencing (e.g. Zoom, Google Meet, Microsoft Teams, Skype, etc.})]

\noindent \textbf{Q.15:} Overall, how often do you share your \textbf{microphone} for all or part of the meeting when you work from home and communicate with others via conferencing tools such as Zoom or Microsoft Teams?
\begin{itemize}
    \item All of my meetings 
    \item Most of my meetings 
    \item About half of my meetings 
    \item A few of my meetings 
    \item None of my meetings 
    \item Other (please specify): [open-text entry]
\end{itemize}

\noindent \textbf{Q.16:} Overall, how often do you share your \textbf{camera} for all or part of the meeting when you work from home and communicate with others via conferencing tools such as Zoom or Microsoft Teams?
\begin{itemize}
    \item All of my meetings 
    \item Most of my meetings 
    \item About half of my meetings 
    \item A few of my meetings 
    \item None of my meetings 
    \item Other (please specify): [open-text entry]
\end{itemize}    

\noindent \textbf{Q.17:} Overall, how often do you share your \textbf{screen} for all or part of the meeting when you work from home and communicate with others via conferencing tools such as Zoom or Microsoft Teams?
\begin{itemize}
    \item All of my meetings 
    \item Most of my meetings 
    \item About half of my meetings 
    \item A few of my meetings 
    \item None of my meetings 
    \item Other (please specify): [open-text entry]
\end{itemize}    

\noindent [\textbf{Q.18} is displayed if the following screening criteria are met: \newline 
\textbf{Q.1} answer is (\quotes{Employee (full-time)} OR \quotes{Employee (part-time)} OR \quotes{Self-employed / Freelancer / Business-owner}) \newline 
AND if \textbf{Q.3} answer is \quotes{No} \newline 
AND if \textbf{Q.6} answer is NOT \quotes{Other (please specify)} \newline 
AND if \textbf{Q.7} answer is NOT \quotes{Other (please specify)} \newline 
AND if \textbf{Q.8} answer is \quotes{Yes} \newline 
AND if \textbf{Q.9} answer is NOT \quotes{Other (please specify)} \newline 
AND if \textbf{Q.10} answer is NOT (\quotes{Less than 3 months} OR \quotes{Other (please specify)}) \newline 
AND if \textbf{Q.11} answer is (\quotes{At least once every work from home day} OR \quotes{At least once every few work from home days} OR \quotes{At least once every work from home week} OR \quotes{At least once every few work from home weeks}) \newline
AND if \textbf{Q.12} answer is NOT \quotes{Other (please specify)} \newline 
AND if \textbf{Q.13} answer is\quotes{Mobile smartphone} OR \quotes{Tablet} OR \quotes{Laptop} OR \quotes{Personal Computer (PC)}\newline 
AND if \textbf{Q.14} answer is \quotes{Video conferencing (e.g. Zoom, Google Meet, Microsoft Teams, Skype, etc.)} OR (\quotes{Phone} OR \quotes{Email messaging} OR \quotes{Instant messaging applications (Slack, WhatsApp, etc.)} OR \quotes{Other (please specify)})]

\noindent \textbf{Q.18:} Do you agree to take a survey about your work from home experience, which will take around \textbf{15 min}? You will be paid additional \textbf{\$3.75} through your Prolific account for taking it.
\begin{itemize}
    \item Yes 
    \item No 
\end{itemize} 

\noindent[\textbf{Q.19} is displayed if \textbf{Q.18} answer is \quotes{Yes}] 

\noindent \textbf{Q.19:} Before we move to the study's main survey, do you have any comments you want to add about the previous screening survey? (optional) \newline 
\noindent [Open-text entry]

\noindent [Redirect to the main survey] 

\subsection{MAIN SURVEY} \label{app:main_survey}
\subsubsection{INTRODUCTION}~\\
\textbf{Work From Home (WFH)} is a flexible work setting in which employees perform their duties including collaborative work with others, such as meeting, teaching, etc. \textbf{remotely from their own homes, mostly using information and communication technologies} (e.g. Zoom, Microsoft Teams, Skype, etc.), instead of from a dedicated office space provided by employers. \newline  

\noindent Most people all over the world have experienced work from home in the last three years due to the COVID-19 pandemic. \newline 

\noindent In this survey, we will ask you a set of multiple choice and open-ended questions about your experience in working from home, in addition to some demographic questions. \newline

\noindent \textcolor{red}{\textbf{Please note that you cannot return back to previous sections once you click the "Next" button. So make sure you selected the answer you really want before you click the \quotes{Next} button.}}  

\subsubsection{WFH SETUP}~\\
\noindent \textbf{Q.1:} How many bedrooms in your home?
\begin{itemize}
    \item 1 bedroom 
    \item 2 bedrooms 
    \item 3 bedrooms 
    \item 4 bedrooms 
    \item 5 bedrooms 
    \item More than 5 bedrooms 
    \item Other (please specify): [open-text entry] 
\end{itemize}

\noindent \textbf{Q.2:} Who lives in your home with you? (select all that apply) 
\begin{itemize}
    \item Spouse or partner
    \item Roommate(s) 
    \item Child(ren) under the age of 18
    \item Child(ren) over the age of 18 
    \item Parent(s) 
    \item Sibling(s) 
    \item No one [exclusive] 
    \item Other (please specify): [open-text entry] 
\end{itemize}

\noindent \textbf{Q.3:} Approximately, how long have you been living in your home described above? 
\begin{itemize}
    \item Less than a year
    \item One year 
    \item 2 years 
    \item 3 years 
    \item 4 years
    \item 5 years 
    \item More than 5 years 
    \item Other (please specify): [open-text entry] 
\end{itemize}

\noindent \textbf{Q.4:} Do you have a dedicated space that you use to work from home? 
\begin{itemize}
    \item Yes
    \item No 
\end{itemize}

\noindent \textbf{Q.5:} Whether you have a dedicated space or not, describe the space that you use to work from home (e.g. a dedicated office room, a corner in the living room, a corner in the bed room, etc.). \newline 
\noindent [Open-text entry]

\subsubsection{WFH SCENARIOS}~\\
In the following sections, we will present you with several scenarios that can happen in work from home settings. For each scenario, we will ask you about your experience.

\noindent [\textbf{Q.6}, \textbf{Q.9}, \textbf{Q.12}, and \textbf{Q.15} are displayed as matrices in randomized-order (the matrices), one matrix per page.]\newline 
\noindent[The following list represents the matrices columns for \textbf{Q.6}, \textbf{Q.9}, \textbf{Q.12}, and \textbf{Q.15}]
\begin{itemize}
    \item Experienced it and felt very uncomfortable
    \item Experienced it and felt somewhat uncomfortable
    \item Experienced it and felt comfortable
    \item Never experienced it
\end{itemize}

\noindent[\textbf{Q.6}, \textbf{Q.9}, \textbf{Q.12}, and \textbf{Q.15} scenario lists represents the matrices rows]\newline

\noindent \textbf{Voice/Sound Scenarios} \newline 
\noindent \textbf{Q.6:} For each scenario, if you experienced the scenario one or more times while working from home \textbf{within the last 3 years}, select how comfortable or uncomfortable you have felt overall. If you never experienced a scenario, select \quotes{Never experienced it.}
\begin{itemize} 
    \item \textbf{Your voice} was inadvertently picked up by your device’s microphone, and was heard by one or more people on your work-related remote call or meeting.
    \item \textbf{The voice of an adult person (other than you) in your home} was inadvertently picked up by your device’s microphone, and was heard by one or more people on your work-related remote call or meeting. 
    \item \textbf{The voice of a child in your home} was inadvertently picked up by your device’s microphone, and was heard by one or more people on your work-related remote call or meeting.
    \item \textbf{The sound of an object (e.g. vacuum cleaner, doorbell, etc.) in your home} was inadvertently picked up by your device’s microphone, and was heard by one or more people on your work-related remote call or meeting. 
    \item \textbf{The sound of a pet in your home} was inadvertently picked up by your device’s microphone, and was heard by one or more people on your work-related remote call or meeting.
\end{itemize}
\noindent [Q.7 is displayed if one or more of \textbf{Q.6} answers were \quotes{Experienced it and felt very uncomfortable} OR \quotes{Experienced it and felt somewhat uncomfortable} OR \quotes{Experienced it and felt comfortable}] \newline 
\noindent \textbf{Q.7:} Has experiencing any of the \textbf{voice/sound scenarios} presented to you in the previous question caused you to adopt protective measures to prevent those incidents from happening again?
\begin{itemize} 
    \item Yes (what are they?) \newline 
    \noindent \textit{Please describe any protective measures you took, including any new software you started using, changes to software settings, changes to your physical workspace, or anything else.} \newline 
    \noindent [Open-text entry] 
    
    \item No (what are the reasons to not adopt protective measures?) \newline 
    \noindent [Open-text entry] 
\end{itemize}

\noindent [Q.8 is displayed if all \textbf{Q.6} answers were \quotes{Never experienced it}] \newline 
\noindent \textbf{Q.8:} Have you adopted protective measures to prevent the \textbf{voice/sound scenarios} presented to you in the previous question from happening to you? 
\begin{itemize} 
    \item Yes (what are they?) \newline 
    \noindent \textit{Please describe any protective measures you took, including any new software you started using, changes to software settings, changes to your physical workspace, or anything else.} \newline
    \noindent [Open-text entry] 
    \item No (what are the reasons to not adopt protective measures?)\newline 
    \noindent [Open-text entry]
\end{itemize}
\noindent\textbf{Video Footage Scenarios} \newline 
\noindent \textbf{Q.9:} For each scenario, if you experienced the scenario one or more times while working from home \textbf{within the last 3 years}, select how comfortable or uncomfortable you have felt overall. If you never experienced a scenario, select \quotes{Never experienced it.}
\begin{itemize} 
    \item \textbf{Video footage of you} was inadvertently captured by your device’s camera, and was seen by one or more people on your work-related remote call or meeting. 
    \item \textbf{Video footage of an adult person (other than you) in your home} was inadvertently captured by your device’s camera, and was seen by one or more people on your work-related remote call or meeting.
    \item \textbf{Video footage of a child in your home} was inadvertently captured by your device’s camera, and was seen by one or more people on your work-related remote call or meeting. 
    \item \textbf{Video footage of an object (e.g. books, furniture, artwork, etc.) in your home} was inadvertently captured by your device’s camera, and was seen by one or more people on your work-related remote call or meeting.
    \item \textbf{Video footage of a pet in your home} was inadvertently captured by your device’s camera, and was seen by one or more people on your work-related remote call or meeting.
\end{itemize}
\noindent [Q.10 is displayed if one or more of \textbf{Q.9} answers were \quotes{Experienced it and felt very uncomfortable} OR \quotes{Experienced it and felt somewhat uncomfortable} OR \quotes{Experienced it and felt comfortable}] \newline 
\noindent \textbf{Q.10:} Has experiencing any of the \textbf{video footage scenarios} presented to you in the previous question caused you to adopt protective measures to prevent those incidents from happening again?
\begin{itemize} 
    \item Yes (what are they?) \newline 
    \noindent \textit{Please describe any protective measures you took, including any new software you started using, changes to software settings, changes to your physical workspace, or anything else.} \newline 
    \noindent [Open-text entry] 
    
    \item No (what are the reasons to not adopt protective measures?) \newline 
    \noindent [Open-text entry] 
\end{itemize}

\noindent [Q.11 is displayed if all \textbf{Q.9} answers were \quotes{Never experienced it}] \newline 
\noindent \textbf{Q.11:} Have you adopted protective measures to prevent the \textbf{video footage scenarios} presented to you in the previous question from happening to you? 
\begin{itemize} 
    \item Yes (what are they?) \newline 
    \noindent \textit{Please describe any protective measures you took, including any new software you started using, changes to software settings, changes to your physical workspace, or anything else.} \newline 
    \noindent [Open-text entry] 
    \item No (what are the reasons to not adopt protective measures?)\newline 
    \noindent [Open-text entry]
\end{itemize}

\noindent \textbf{Computer/Browsing Data Scenarios} \newline 
\noindent \textbf{Q.12:} For each scenario, if you experienced the scenario one or more times while working from home \textbf{within the last 3 years}, select how comfortable or uncomfortable you have felt overall. If you never experienced a scenario, select \quotes{Never experienced it.}
\begin{itemize} 
    \item \textbf{Your data (e.g. emails, files, images, file names, etc.)} was inadvertently displayed on your device’s shared screen, and was seen by one or more people on your work-related remote call or meeting.
    \item \textbf{Your personalized web browsing data (e.g. personalized ads, auto-completed forms/URLs, browser opened tabs, etc.)} was inadvertently displayed on your device’s shared screen, and was seen by one or more people on your work-related remote call or meeting.
\end{itemize}
\noindent [Q.13 is displayed if one or more of \textbf{Q.12} answers were \quotes{Experienced it and felt very uncomfortable} OR \quotes{Experienced it and felt somewhat uncomfortable} OR \quotes{Experienced it and felt comfortable}] \newline 
\noindent \textbf{Q.13:} Has experiencing any of the \textbf{computer/browsing data scenarios} presented to you in the previous question caused you to adopt protective measures to prevent those incidents from happening again?
\begin{itemize} 
    \item Yes (what are they?) \newline 
    \noindent \textit{Please describe any protective measures you took, including any new software you started using, changes to software settings, changes to your physical workspace, or anything else.} \newline 
    \noindent [Open-text entry] 
    
    \item No (what are the reasons to not adopt protective measures?) \newline 
    \noindent [Open-text entry] 
\end{itemize}

\noindent [Q.14 is displayed if all \textbf{Q.12} answers were \quotes{Never experienced it}] \newline 
\noindent \textbf{Q.14:} Have you adopted protective measures to prevent the \textbf{computer/browsing data scenarios} presented to you in the previous question from happening to you? 
\begin{itemize} 
    \item Yes (what are they?) \newline 
    \noindent \textit{Please describe any protective measures you took, including any new software you started using, changes to software settings, changes to your physical workspace, or anything else.} \newline 
    \noindent [Open-text entry] 
    \item No (what are the reasons to not adopt protective measures?)\newline 
    \noindent [Open-text entry]
\end{itemize}

\noindent \textbf{Autonomy Scenarios} \newline 
\noindent \textbf{Q.15:} For each scenario, if you experienced the scenario one or more times while working from home \textbf{within the last 3 years}, select how comfortable or uncomfortable you have felt overall. If you never experienced a scenario, select \quotes{Never experienced it.}
\begin{itemize} 
   \item \textbf{You} wanted to do something urgent privately (e.g. take medicine) at your home, but were not allowed to stop sharing the camera.
   \item \textbf{You} wanted to have an urgent private conversation at your home, but were not allowed to stop sharing the microphone.
\end{itemize}
\noindent [Q.16 is displayed if one or more of \textbf{Q.15} answers were \quotes{Experienced it and felt very uncomfortable} OR \quotes{Experienced it and felt somewhat uncomfortable} OR \quotes{Experienced it and felt comfortable}] \newline 
\noindent \textbf{Q.16:} Has experiencing any of the \textbf{autonomy scenarios} presented to you in the previous question caused you to adopt protective measures to prevent those incidents from happening again?
\begin{itemize} 
    \item Yes (what are they?) \newline 
    \noindent \textit{Please describe any protective measures you took, including any new software you started using, changes to software settings, changes to your physical workspace, or anything else.} \newline 
    \noindent [Open-text entry] 
    
    \item No (what are the reasons to not adopt protective measures?) \newline 
    \noindent [Open-text entry] 
\end{itemize}

\noindent [Q.17 is displayed if all \textbf{Q.15} answers were \quotes{Never experienced it}] \newline 
\noindent \textbf{Q.17:} Have you adopted protective measures to prevent the \textbf{autonomy scenarios} presented to you in the previous question from happening to you? 
\begin{itemize} 
    \item Yes (what are they?) \newline 
    \noindent \textit{Please describe any protective measures you took, including any new software you started using, changes to software settings, changes to your physical workspace, or anything else.} \newline 
    \noindent [Open-text entry] 
    \item No (what are the reasons to not adopt protective measures?)\newline 
    \noindent [Open-text entry]
\end{itemize}
\subsubsection{DETAILED SCENARIO}~\\
\noindent \textbf{Q.18:} From the list, please select your \textbf{most memorable incident} that you reported in the previous sections when you felt very or somewhat uncomfortable. \newline
[A list of the scenarios that the participant reported that they made them feel \quotes{very uncomfortable} or \quotes{somewhat uncomfortable} in Q.6, Q.9, Q.12, and Q.15 is displayed here] \newline 
[\textbf{Q.19 - Q.30} are displayed if the participant selected one ore more scenario as \quotes{very uncomfortable} or \quotes{somewhat uncomfortable} in \textbf{Q.6}, \textbf{Q.9}, \textbf{Q.12}, and \textbf{Q.15}]\newline 
\noindent \textbf{Q.19:} In your own words, describe in more detail this specific incident in which you felt very or somewhat uncomfortable: \newline 
\noindent[Open-text entry]

\noindent \textbf{Q.20:} How frequently has this incident happened to you?
\begin{itemize} 
   \item Once
   \item A few times 
   \item Repeatedly
   \item I cannot remember 
\end{itemize}

\noindent \textbf{Q.21:} Why did it make you feel uncomfortable? \newline 
\noindent [Open-text entry]
   
\noindent \textbf{Q.22:} To what extent did the discomfort escalate to cause harm (e.g. financial, physical, psychological, etc.) to yourself or others?
\begin{itemize} 
   \item It caused a large amount of harm
   \item It caused a moderate amount of harm
   \item It caused a small amount of harm 
   \item It did not cause any harm
\end{itemize}

\noindent[\textbf{Q.23 - Q.24} are displayed if \textbf{Q.22} answer is \quotes{It caused a large amount of harm} OR \quotes{It caused a moderate amount of harm} OR \quotes{It caused a small amount of harm}]\newline 
\noindent \textbf{Q.23:} What type of harm did it cause? (select all that apply).
\begin{itemize} 
   \item Financial harm
   \item Physical harm
   \item Psychological harm 
   \item Other (please specify): [open-text entry] 
\end{itemize}

\noindent \textbf{Q.24:} Please explain how it caused this harm? \newline 
\noindent [Open-text entry]

\noindent[\textbf{Q.25} is displayed if \textbf{Q.22} answer is \quotes{It did not cause any harm}]\newline
\noindent \textbf{Q.25:} Please explain how and why it did not cause any harm? \newline 
\noindent [Open-text entry]

\noindent \textbf{Q.26:} Where in your home were you located at the time of the incident you described?
\begin{itemize} 
   \item Office room
   \item Living room / recreation area 
   \item Bedroom 
   \item Kitchen 
   \item Dining room 
   \item Garage 
   \item Bathroom 
   \item Other (please specify): [open-text entry] 
\end{itemize}

\noindent \textbf{Q.27:} Which of the following were you sharing during the incident you described? (select all that apply)
\begin{itemize} 
    \item Microphone 
    \item Camera 
    \item Screen 
    \item I cannot remember [exclusive] 
    \item Other (*please specify): [open-text entry]
\end{itemize}

\noindent \textbf{Q.28:} Was the incident recorded?
\begin{itemize} 
    \item Yes, video and voice
    \item Yes, video only
    \item Yes, voice only
    \item No, not recorded
    \item I cannot remember 
\end{itemize}

\noindent \textbf{Q.29:} What was your relationship with the people at work with whom this incident happened? (e.g. colleagues, managers, customers, students, etc.)? \newline 
\noindent[Open-text entry]

\noindent \textbf{Q.30:} How did this relationship with the people at work affect the impact of the incident on you? \newline 
\noindent[Open-text entry]

\noindent[\textbf{Q.31 - Q.32} are displayed if the participants did not select any scenario as \quotes{very uncomfortable} or \quotes{somewhat uncomfortable} in \textbf{Q.6}, \textbf{Q.9}, \textbf{Q.12}, and \textbf{Q.15}] \newline 
\noindent \textbf{Q.31:} In your own words, describe in more detail one incident that, \textbf{if you WERE to experience it while working from home}, would make you feel very or somewhat uncomfortable? \newline 
\noindent[Open-text entry]   

\noindent \textbf{Q.32:} Why would it make you feel uncomfortable? \newline 
\noindent[Open-text entry]  

\subsubsection{Demographics}~\\
\noindent Please answer the following demographic questions.

\noindent \textbf{Q.33:} Which of the following technologies, if any, does your employer use to monitor your productivity when you work from home?  (select all that apply) 
\begin{itemize} 
    \item Time trackers 
    \item Activity trackers 
    \item Task trackers
    \item Video monitoring
    \item Audio monitoring
    \item I am not aware of any productivity trackers used by my employers [exclusive]
    \item Other (*please specify): [open-text entry]
\end{itemize}
\noindent \textbf{Q.34:} What is your age in years?
\begin{itemize}
    \item From 18 to 24
    \item From 25 to 34 
    \item From 35 to 44 
    \item From 45 to 54 
    \item From 55 to 64
    \item From 65 to 74
    \item 75 or older
    \item Prefer not to answer
\end{itemize}

\noindent \textbf{Q.35:} What is your gender?
\begin{itemize}
    \item Male
    \item Female 
    \item Non-binary
    \item Prefer to self describe: [open-text entry] 
    \item Prefer not to answer 
\end{itemize}

\noindent \textbf{Q.36:} What is your race or ethnic identity? (You may select more than one option)
\begin{itemize}
    \item White 
    \item Black or African American 
    \item American Indian or Alaska Native 
    \item Asian 
    \item Native Hawaiian or Pacific Islander
    \item Hispanic and/or Latino/Latina/Latinx
    \item Prefer to self describe: [open-text entry]
    \item Prefer not to answer [exclusive] 
\end{itemize}

\noindent \textbf{Q.37:} What is your approximate annual household income? \newline 
\noindent \textit{Please answer based on your entire current annual household's income, before taxes.}
\begin{itemize}
    \item Less than \$20,000 
    \item \$20,000 to \$39,999 
    \item \$40,000 to \$59,999
    \item \$60,000 to \$79,999
    \item \$80,000 to \$99,999
    \item \$100,000 to \$149,999
    \item \$150,000 or more 
    \item Prefer not to answer 
\end{itemize}

\noindent \textbf{Q.38:} What is the highest educational degree you have received?
\begin{itemize}
    \item Doctoral degree
    \item Master's degree
    \item Bachelor's degree
    \item Associate's degree
    \item High school diploma or GED
    \item Less than high school degree
    \item Other (please specify): [open-text entry]
\end{itemize}

\noindent \textbf{Q.39:} Do you have a university degree in, or currently work in, one or more of the following fields: Computer Science (CS), Information Systems (IS), Information Technology (IT), or Computer Engineering (CE)?
\begin{itemize}
    \item Yes
    \item No
\end{itemize}

\noindent \textbf{Q.40:} Do  you have a university degree in cybersecurity or currently work in the cybersecurity area?
\begin{itemize}
    \item Yes
    \item No
\end{itemize}

\noindent \textbf{Q.41:} If you have any other thoughts or feedback about this survey or the information you viewed, please let us know here. (optional) \newline 
\noindent[Open-text entry]

\noindent \textbf{You now reached the end of the survey. To submit your response click the \quotes{Submit} button.}
\clearpage
\section{Appendix B - Further Results} \label{app:more_results}
\subsection{Screening Survey Results} \label{sec:screening}
\autoref{tab:wfh_context} and \autoref{tab:conf_tools_usage} list the screening survey results for our participants.
\begin{table}[h!]
\centering
\setcounter{table}{4}  
\renewcommand*{\arraystretch}{1.25}
\caption{Participants' work context.}
\label{tab:wfh_context}
\resizebox{0.77\columnwidth}{!}{
\begin{tabular}{p{6cm}|>{}l|>{}l}
\toprule
\multicolumn{3}{c}{\textit{N = 214}} \\
    \midrule
    Employment type            & No.   & \% \\
    \midrule
    Employee (full-time)     & 175   & 81.8\% \\ 
    Employee (part-time)	     & 19	   & 8.9\% \\ 
    Self-employed / Freelancer / Business-owner & 20	& 9.3\% \\
    \midrule 
    Work sector	             & No.	& \% \\
    \midrule
    Pre-university Education	& 3	 & 1.4\% \\
    University Education	    & 11 & 5.1\% \\
    Health	                & 23 & 10.7\% \\
    Information and Communication Technology	& 48 & 22.4\% \\
    Financial	                & 29	& 13.6\% \\
    Industrial	            & 7	    & 3.3\% \\
    Agricultural	            & 0	   & 0.0\% \\
    Sales and retail        & 26   & 12.1\% \\
    Petrochemical	            & 0	   & 0.0\% \\
    Other (please specify)	& 67   & 31.3\% \\ 
    \midrule 
    Organization type         & No.	& \% \\ 
    \midrule 
    Privately-held organization & 116	& 54.2\% \\
    Publicly-traded organization   & 41	& 19.2\% \\
    Government organization	     & 25	& 11.7\% \\
    Educational organization	     & 14	& 6.5\% \\
    Not-for-profit organization	 & 17	& 7.9\% \\
    Other (please specify)	     & 1	& 0.5\% \\
    \midrule 
    Duration working in current job	& No.	& \% \\ 
    \midrule 
    Less than a year	                & 16	& 7.5\% \\
    One year	                        & 23	& 10.7\% \\
    2 years	                        & 42	& 19.6\% \\
    3 years	                        & 20	& 9.3\% \\
    4 years	                        & 14	& 6.5\% \\
    5 years	                        & 18	& 8.4\% \\
    More than 5 years	                & 81	& 37.9\% \\
    Other (please specify)	        & 0	    & 0.0\% \\
    \midrule
    Avg. normal work days/week	    & No.	& \%    \\
    \midrule 
    One day per week	                & 1	    & 0.5\% \\
    2 days per week	                & 3	    & 1.4\% \\
    3 days per week	                & 7	    & 3.3\% \\
    4 days per week	                & 11	& 5.1\% \\
    5 days per week	                & 177	& 82.7\% \\
    6 days per week	                & 12	& 5.6\% \\
    7 days per week	                & 3	    & 1.4\% \\
   Other (please specify)	         & 0	    & 0.0\% \\
    \midrule 
     Avg. WFH days/week	           & No.	& \%    \\
    \midrule 
    One day per week	                & 18	& 8.4\%  \\
    2 days per week	                & 28	& 13.1\% \\
    3 days per week	                & 33	& 15.4\% \\
    4 days per week	                & 31	& 14.5\% \\
    5 days per week	                & 95	& 44.4\% \\
    6 days per week	                & 7	    & 3.3\% \\
    7 days per week	                & 2	    & 0.9\% \\
    Other (please specify)	        & 0	    & 0.0\% \\
    \midrule 
    Duration been WFH                 & No.	& \% \\  
    \midrule
    Less than 3 months	           & 0	    & 0.0\% \\
    From 3 to 11 months	           & 21	    & 9.8\% \\
    One year	                       & 23	    & 10.7\% \\
    2 years	                       & 45	    & 21.0\% \\
    3 years	                       & 66	    & 30.8\% \\
    4 years	                       & 27	    & 12.6\% \\
    5 years	                       & 9	    & 4.2\% \\
    More than 5 years	               & 23	    & 10.7\% \\
    Other (please specify)	       & 0	    & 0.0\% \\
\bottomrule
\end{tabular}
}
\end{table}
\begin{table}[h!]
\centering  
\renewcommand*{\arraystretch}{1.25}
\caption{Participants' conference tools usage.}
\label{tab:conf_tools_usage}
\resizebox{0.77\columnwidth}{!}{
\begin{tabular}{p{7cm}|>{}l|>{}l}
\toprule
\multicolumn{3}{c}{\textit{N = 214}} \\
    \midrule
    Frequency comm. via conference tools    &  No.	& \% \\
    \midrule 
    At least once every work from home day	    & 132	& 61.7\% \\
    At least once every few work from home days   & 43	& 20.1\% \\
    At least once every work from home week	    & 28	& 13.1\% \\
    At least once every few work from home weeks	& 11	& 5.1\% \\
    At least once every work from home month	    & 0	    & 0.0\% \\
    Less than once every work from home month	    & 0	    & 0.0\% \\
    Other (please specify)	                    & 0	    & 0.0\% \\
    \midrule 
    Avg. time spent comm via conference tools & No.	& \% \\  
    \midrule
    Less than one hour per work from home day	& 51 & 23.8\% \\
    At least one hour per work from home day	& 82 &	38.3\% \\
    At least 2 hours per work from home day	    & 46 & 21.5\% \\
    At least 3 hours per work from home day	    & 19 & 8.9\% \\
    At least 4 hours per work from home day	    & 6	 & 2.8\% \\
    More than 4 hours per work from home day	& 10 & 4.7\% \\
    Other (please specify)	                    & 0	 & 0.0\% \\
    \midrule 
    Devices used in WFH (non-exclusive)	    & No.	& \% \\
    \midrule 
    Mobile smartphone	                        & 151	& 70.6\% \\
    Mobile traditional (non-smart) phone	    & 0	    & 0.0\% \\ 
    Landline phone	                            & 13	& 6.1\% \\
    Tablet                                    & 28	& 13.1\% \\
    Laptop	                                    & 194	& 90.7\% \\
    Personal Computer (PC)	                    & 74	& 34.6\% \\
    Other (please specify)	                    & 0	    & 0.0\%\\
    \midrule 
    Techs used in WFH (non-exclusive)	& No.	& \% \\
    \midrule 
    Video conferencing (e.g. Zoom, Google Meet, Microsoft Teams, Skype, etc.)	& 214	& 100.0\% \\
    Phone (audio only)	& 124	& 57.9\% \\
    Email messaging	    & 194	& 90.7\% \\
    Instant messaging applications (Slack, WhatsApp, etc.)	& 141	& 65.9\% \\
    Mobile Short Messaging Service (SMS) messaging        & 79	& 36.9\% \\
    Other (please specify)	                                & 3	    & 1.4\% \\
    \midrule 
    Frequency share microphone	               & No.	& \% \\
    \midrule 
    All of my meetings	                       & 75	    & 35.0\% \\
    Most of my meetings	                       & 74	    & 34.6\% \\
    About half of my meetings	               & 30	    & 14.0\% \\
    A few of my meetings	                   & 31	    & 14.5\% \\
    None of my meetings	                       & 4	    & 1.9\% \\
    \midrule 
    Frequency share camera	                   & No.	& \% \\
    \midrule 
    All of my meetings	                       & 46	    & 21.5\% \\
    Most of my meetings	                       & 56	    & 26.2\% \\
    About half of my meetings	               & 32	    & 15.0\% \\
    A few of my meetings	                   & 62	    & 29.0\% \\
    None of my meetings	                       & 18	    & 8.4\% \\
    \midrule 
    Frequency share screen	                   & No.	& \% \\
    \midrule 
    All of my meetings	                       & 8	    & 3.7\% \\
    Most of my meetings	                       & 20	    & 9.3\% \\
    About half of my meetings	               & 41	    & 19.2\% \\
    A few of my meetings	                   & 121	& 56.5\% \\
    None of my meetings	                       & 24	    & 11.2\% \\
\bottomrule
\end{tabular}
}
\end{table}
\clearpage
\subsection{Main Survey Results} \label{sec:main}
\autoref{tab:demographics} and~\autoref{tab:wfh_setting} list demographic details and WFH setting for our participants.
\begin{table}[h!]
\centering  
\renewcommand*{\arraystretch}{1.25}
\caption{Participants' demographics.}
\label{tab:demographics}
\resizebox{0.77\columnwidth}{!}{
\begin{tabular}{p{7cm}|l|l}
\toprule
\multicolumn{3}{c}{\textit{N = 214}} \\
\midrule
Employer productivity monitoring (non-exclusive)	& No.	& \% \\
\midrule
Time trackers	       & 41	& 19.2\% \\
Activity trackers	   & 39	& 18.2\% \\
Task trackers        & 30	& 14.0\% \\
Video monitoring	    & 11	& 5.1\% \\ 
Audio monitoring	    & 10	& 4.7\% \\ 
I am not aware of any productivity trackers used by my employers	            & 143	& 66.8\% \\ 
Other (*please specify)	& 6	& 2.8\% \\
\midrule 
Age (years) & No. & \% \\
\midrule
From 18 to 24 & 17 & 7.9\% \\
From 25 to 34 & 81 & 37.9\% \\ 
From 35 to 44 & 67 & 31.3\% \\
From 45 to 54 & 25 & 11.7\% \\
From 55 to 64 & 21 & 9.8\% \\
From 65 to 74 & 3 & 1.4\% \\
75 or older & 0 & 0.0\% \\
Prefer not to answer & 0 & 0.0\% \\
\midrule
Gender & No. & \% \\
\midrule
Male & 116 & 54.2\% \\
Female & 94 & 43.9\% \\
Non-binary & 3 & 1.4\% \\
Prefer to self describe: & 0 & 0.0\% \\
Prefer not to answer & 1 & 0.5\% \\
\midrule
Race / Ethnicity (non-exclusive)	 & No. & \% \\
\midrule
White & 154 & 72.0\% \\ 
Black or African American & 23 & 10.7\% \\ 
American Indian or Alaska Native & 7 & 3.3\% \\
Asian & 33 & 15.4\% \\
Native Hawaiian or Pacific Islander & 0 & 0.0\% \\
Hispanic and/or Latino/Latina/Latinx & 16 & 7.5\% \\
Prefer to self describe: & 3 & 1.4\% \\
Prefer not to answer & 3 & 1.4\% \\
\midrule
Annual Income & No. & \% \\
\midrule 
Less than \$20,000 & 3 & 1.4\% \\
\$20,000 to \$39,999 & 13 & 6.1\% \\
\$40,000 to \$59,999 & 31 & 14.5\% \\
\$60,000 to \$79,999 & 49 & 22.9\% \\
\$80,000 to \$99,999 & 33 & 15.4\% \\
\$100,000 to \$149,999 & 43 & 20.1\% \\
\$150,000 or more & 37 & 17.3\% \\
Prefer not to answer & 5 & 2.3\% \\
\midrule
Education & No. & \% \\
\midrule 
Doctoral degree & 6 & 2.8\% \\
Master's degree & 43 & 20.1\% \\
Bachelor's degree & 108 & 50.5\% \\
Associate's degree & 17 & 7.9\% \\
High school diploma or GED & 38 & 17.8\% \\
Less than high school degree & 0 & 0.0\% \\
Other (please specify) & 2 & 0.9\% \\
\midrule
CS/IS/IT/CE Background & No. & \% \\ 
\midrule
Yes & 52 & 24.3\% \\
No & 162 & 75.7\% \\
\midrule 
Cybersecurity Background & No. & \% \\
\midrule
Yes & 16 & 7.5\% \\ 
No & 198 & 92.5\% \\
\bottomrule
\end{tabular}
}
\end{table}
\begin{table}[h!]
\centering
\caption{Participants' WFH setting.}
\label{tab:wfh_setting}
\resizebox{0.77\columnwidth}{!}{
\renewcommand*{\arraystretch}{1.25}
\begin{tabular}{p{6cm}|l|l}
\toprule
\multicolumn{3}{c}{\textit{N = 214}} \\
\midrule
Bedrooms in home & No. & \% \\
\midrule
1 bedroom & 22 & 10.3\% \\
2 bedrooms & 54 & 25.2\% \\
3 bedrooms & 79 & 36.9\% \\
4 bedrooms & 42 & 19.6\% \\
5 bedrooms & 14 & 6.5\% \\
More than 5 bedrooms & 1 & 0.5\% \\
Other (please specify) & 2 & 0.9\% \\
\midrule
Living with (non-exclusive) & No. & \% \\
\midrule
Spouse or partner & 123 & 57.5\% \\
Roommate(s) & 15 & 7.0\% \\
Child(ren) under the age of 18 & 68	& 31.8\% \\
Child(ren) over the age of 18 & 17	& 7.9\% \\ 
Parent(s) & 36 & 16.8\% \\
Sibling(s) & 14 & 6.5\% \\
No one & 31 & 14.5\% \\
Other (please specify) & 6 & 2.8\% \\
\midrule
Duration living in the home & No. & \% \\
\midrule
Less than a year & 16 & 7.5\% \\
One year & 22 & 10.3\% \\
2 years & 30 & 14.0\% \\
3 years & 21 & 9.8\% \\
4 years & 15 & 7.0\% \\
5 years & 6 & 2.8\% \\
More than 5 years & 102 & 47.7\% \\
Other (please specify) & 2 & 0.9\% \\
\midrule
Have a dedicated space for work & No. & \% \\
\midrule
Yes & 170 & 79.4\% \\
No & 44 & 20.6\% \\
\bottomrule
\end{tabular}
} 
\end{table}

\begin{table*}[h!]
\centering
\renewcommand*{\arraystretch}{1.25}
\caption{The privacy invasive scenarios ordered by categories. Columns 2--4 respectively list the number of participants who experienced the scenario and felt very uncomfortable; somewhat uncomfortable; and comfortable. The last column shows the percentage of participants who experienced the scenario and felt uncomfortable with respect to the number of participants who experienced that scenario (ratio of discomfort). Refer to Table 1 in the main paper for description of the scenarios codes.}
\label{tab:prevalence}
\resizebox{0.8\textwidth}{!}{%
\begin{tabular}{l|l|l|lp{0cm}|l||l|l|l}
\toprule
\multicolumn{9}{c}{\textit{N = 214}} \\
\midrule
\multirow{3}{*}{Scenario code} & \multicolumn{3}{c}{Experienced the scenario \& felt:} & & \multirow{3}{*}{\makecell[l]{Never\\experienced}} & \multirow{3}{*}{\makecell[l]{Total uncomf.\\(somewhat or very)}} & \multirow{3}{*}{\makecell[l]{Total experienced\\(comf.\& uncomf.)}} & \multirow{3}{*}{\makecell[l]{Ratio\\discomf.}} \\

 & \makecell[l]{Very\\uncomf.} & \makecell[l]{Somewhat\\uncomf.}  & \makecell[l]{Comf.} &  & & &	\\
\midrule
\rowcolor{gray!30}
Audio Scenarios & \multicolumn{8}{c}{}\\
\midrule
voice\_you	             & 17 & 44  & 39 & & 114    &  61  & 100   & 61.0\% \\
\hline
voice\_adult	         & 12 & 35  & 50 & & 117     &  47 & 97   & 48.5\% \\
\hline
voice\_child	         & 5 & 20   & 33 & & 156    &  25  & 58    & 43.1\% \\
\hline
sound\_object            & 10 & 45  & 85 & & 74     &  55  & 140   & 39.3\% \\
\hline
sound\_pet	             & 9 & 42   & 71 & & 92     &  51  & 122   & 41.8\% \\
\midrule
\rowcolor{gray!30}
Video Scenarios & \multicolumn{8}{c}{} \\
\midrule
video\_you	             & 8 & 23  & 23  & & 160    & 31   &  54   & 57.4\% \\
\hline
video\_adult	         & 5 & 15  & 33  & & 161    & 20   &  53   & 37.7\% \\
\hline
video\_child	         & 6 & 8   & 25  & & 175    & 14   &  39   & 35.9\% \\
\hline
video\_object	         & 1 & 21  & 92  & & 100    & 22   &  114  & 19.3\% \\
\hline
video\_pet	             & 0 & 11  & 69  & & 134    & 11   &  80   & 13.8\% \\
\midrule
\rowcolor{gray!30}
Data Scenarios & \multicolumn{8}{c}{} \\
\midrule
computer\_data	         & 6 & 25  & 46  & & 137    & 31   &  77   & 40.3\% \\
\hline
browsing\_data	         & 4 & 30  & 35  & & 145    & 34   &  69   & 49.3\% \\
\midrule 
\rowcolor{gray!30}
Autonomy Scenarios & \multicolumn{8}{c}{} \\
\midrule
cannot\_stop\_camera	 & 9 & 33  & 15  & & 157    & 42   &  57   & 73.7\% \\
\hline
cannot\_stop\_mic	     & 11 & 26 & 19  & & 158    & 37   &  56   & 66.1\% \\
\hline 
\end{tabular}
}
\end{table*}

\begin{table}[h!]
\centering
\footnotesize
\caption{The aggregated number of participants who experienced at least one scenario of the named category and felt comfortable or uncomfortable (experienced) and those who felt uncomfortable (very or somewhat).}
\label{tab:cat_agg_summary}
\begin{tabular}{l|l|l}
\toprule 
\multicolumn{3}{c}{\textit{N = 214}} \\
\midrule
Scenario cat. &	Experienced & Uncomf. \\
\midrule 
Audio	 &   187	& 107 \\
\hline 
Video	 &   138	& 55 \\
\hline 
Data	 &   91	    & 54 \\
\hline 
Autonomy &	 72     & 47 \\
\bottomrule
\end{tabular}
\end{table}

\begin{table}[h!]
\centering
\footnotesize
\renewcommand*{\arraystretch}{1.25}
\caption{Categories of the most memorable scenario the participants experienced and made them feel uncomfortable.}
\label{tab:memorable_scenarios}
\begin{tabular}{l|l|l}
\toprule
\multicolumn{3}{c}{\textit{N = 140}} \\
\midrule
Scenario & No. & \% \\
\midrule
sound\_pet & 28 & 20.0\% \\
voice\_you & 19 & 13.6\% \\
voice\_adult &  14 & 10.0\% \\
sound\_object & 	13 & 9.3\% \\
cant\_stop\_camera & 	12 & 8.6\% \\
video\_you	& 11 &	7.9\% \\
computer\_data &	9 &	6.4\% \\
browsing\_data &	7 &	5.0\% \\
voice\_child &	7 &	5.0\% \\
cant\_stop\_mic &	7 &	5.0\% \\
video\_adult &	5 &	3.6\% \\
video\_child &	4 &	2.9\% \\
video\_object &	2& 	1.4\% \\ 
video\_pet &	2	& 1.4\% \\ 
\bottomrule
\end{tabular}
\end{table}

\begin{table}[h!]
\centering
\caption{Characteristics of scenarios that caused discomfort.}
\label{tab:detailed}
\resizebox{0.77\columnwidth}{!}{
\renewcommand*{\arraystretch}{1.25}
\begin{tabular}{p{6cm}|l|l}
\toprule
\multicolumn{3}{c}{\textit{N = 140}} \\
\midrule
Discomfort escalated to cause harm	& No.    & \%  \\
\midrule
It caused a large amount of harm    & 1     &   0.7\%  \\ 
It caused a moderate amount of harm & 0     &   0.0\%  \\ 
It caused a small amount of harm    & 35    &   25.0\%  \\ 
It did not cause any harm           & 104   &   74.3\%  \\ 
\midrule
Type of harm (non-exclusive; \textit{N = 36})	& No. & \%  \\
\midrule
Financial harm	                    & 1     &   2.8\%  \\ 
Physical harm	                    & 1     &   2.8\%  \\ 
Psychological harm	                & 34    &   94.4\% \\ 
Other (please specify)	            & 1     &   2.8\%  \\ 
\midrule
Frequency of event	                & No.     & \% \\
\midrule
Once	                            & 71    &   50.7\% \\ 
A few times	                        & 66    &   47.1\% \\ 
Repeatedly	                        & 3     &   2.1\%  \\ 
I cannot remember	                & 0     &   0.0\%  \\  
\midrule
Recording of event	                & No. & \% \\
\midrule
Yes, video and voice	            & 25    & 17.9\%   \\	
Yes, video only	                    & 4	    & 2.9\%    \\
Yes, voice only	                    & 15	& 10.7\%   \\
No, not recorded	                & 88    & 62.9\%   \\	
I cannot remember	                & 8     & 5.7\%    \\
\midrule
Where in the house	                & No. & \% \\
\midrule
Office room	                        & 59    &  42.1\%  \\	
Living room / recreation area	    & 26	&  18.6\%  \\
Bedroom		                        & 30    &  21.4\%  \\
Kitchen	                            & 7     &  5.0\%  \\	
Dining room	                        & 8     &  5.7\%  \\	
Garage	                            & 0     &  0.0\% \\	
Bathroom	                        & 1	    &  0.7\% \\
Other (please specify)      	    & 9     &  6.4\% \\	
\midrule
Device shared (non-exclusive)	    & No. & \% \\
\midrule
Microphone	                        & 115   &  82.1\% \\	
Camera	                            & 82    &  58.6\% \\	
Screen	                            & 28	&  20.0\% \\
I cannot remember	                & 1	    &  0.7\% \\
Other (please specify)     	        & 2     &  1.4\% \\	
\bottomrule
\end{tabular}
} 
\end{table}

\clearpage
\begin{figure*}[h!]
\centering
\setcounter{figure}{4}  
\includegraphics[width=0.7\textwidth]{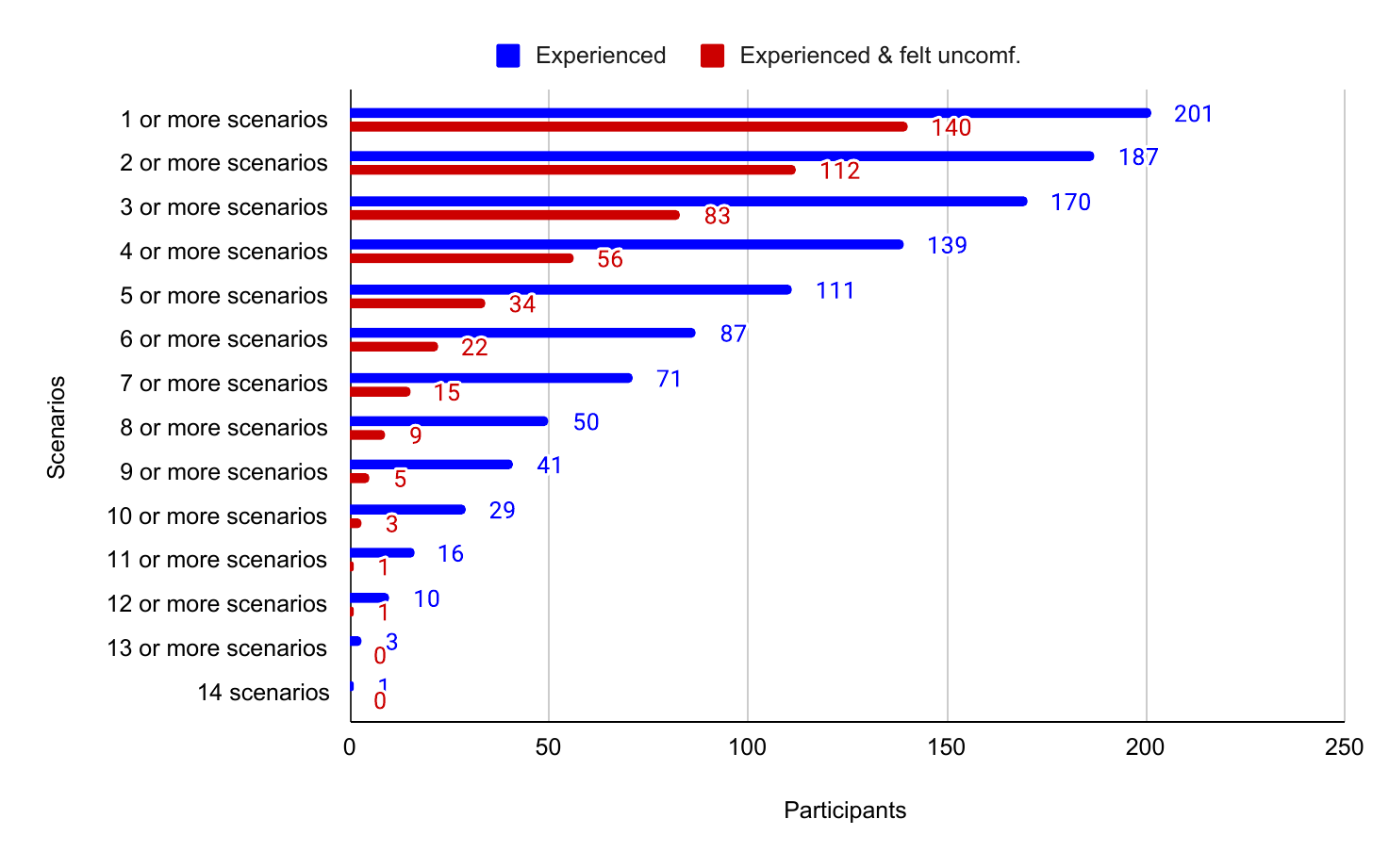}
\caption{Number of participants who experienced one or more scenario (whether they made them feel comfortable or uncomfortable) compared to those who experienced them and felt uncomfortable.}
\label{fig:prevalence}
\end{figure*}

\begin{figure*}[h!]
\centering
\includegraphics[width=0.7\textwidth]{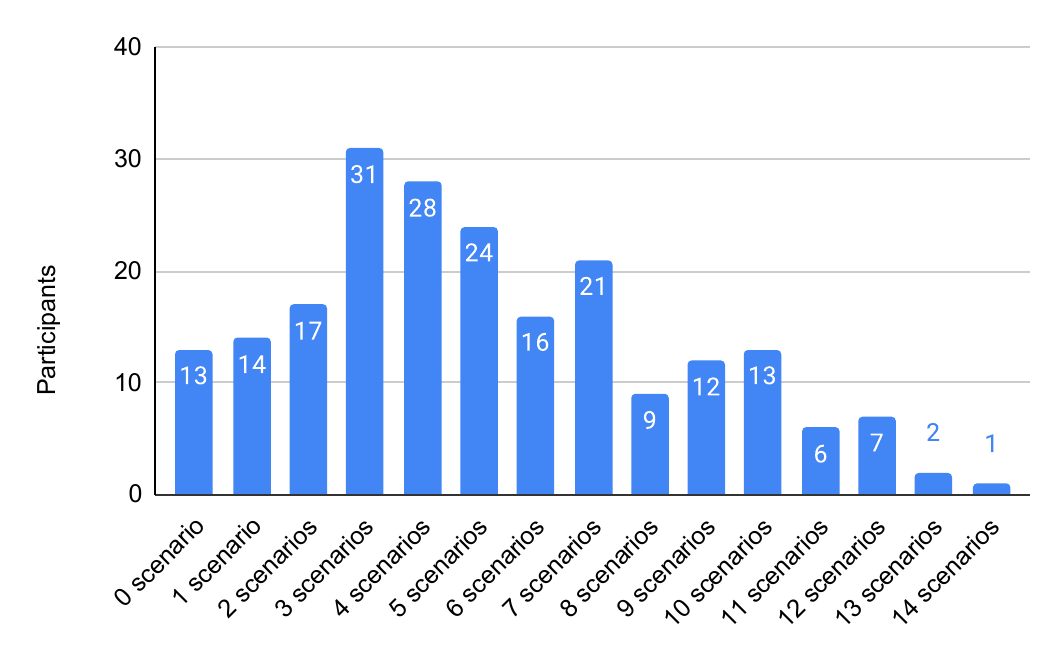}
\caption{Distribution of participants according to the number of scenarios they experienced (whether they made them feel comfortable or uncomfortable).}
\label{fig:freq_exp}
\end{figure*}


\begin{figure*}[h!]
\centering
\includegraphics[width=0.7\textwidth]{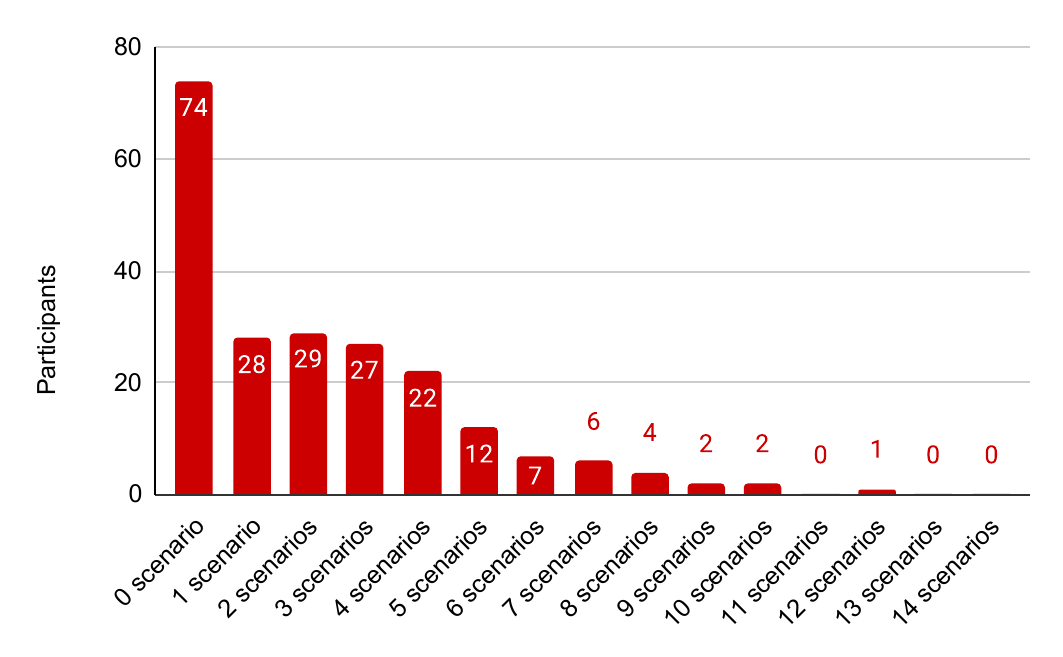}
\caption{Distribution of participants according to the number of scenarios they experienced and made them feel uncomfortable.}
\label{fig:freq_uncomf}
\end{figure*}

\begin{figure*}[h!]
\centering
\includegraphics[width=0.7\textwidth]{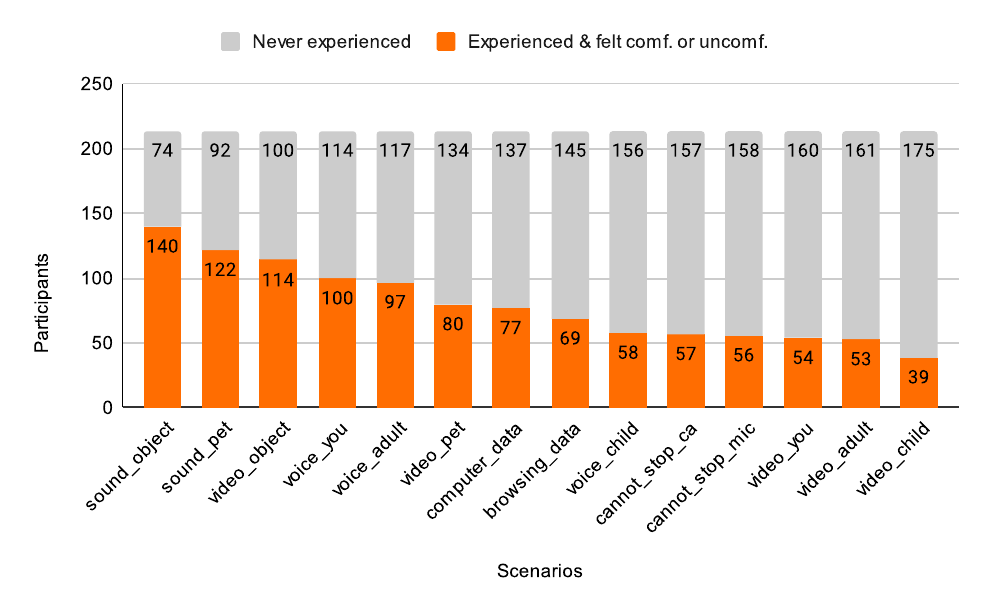}
\caption{Number of participants who experienced the named scenario (whether they made them feel comfortable or uncomfortable) with respect to the total number of participants (whether they experienced the scenario or never did).}
\label{fig:scenario_prevalence_dissected}
\end{figure*}

\begin{figure*}[h!]
\centering
\includegraphics[width=0.7\textwidth]{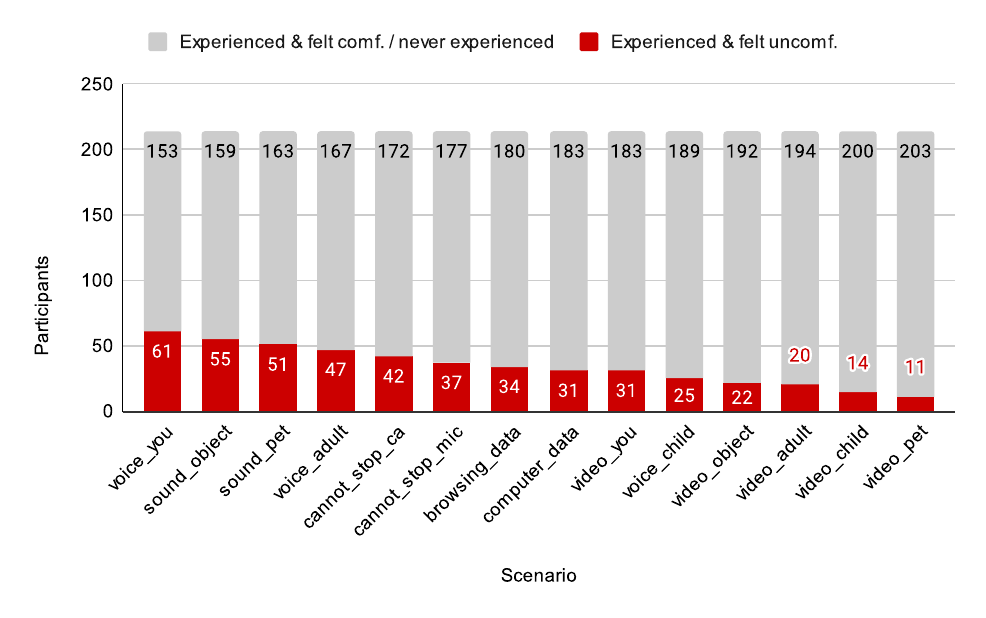}
\caption{Number of participants who experienced the named scenario and felt uncomfortable with respect to the total number of participants (whether they experienced the scenario or never did).}
\label{fig:discomfort_prevalence_dissected}
\end{figure*}

\begin{figure*}[h!]
\centering
\includegraphics[width=0.7\textwidth]{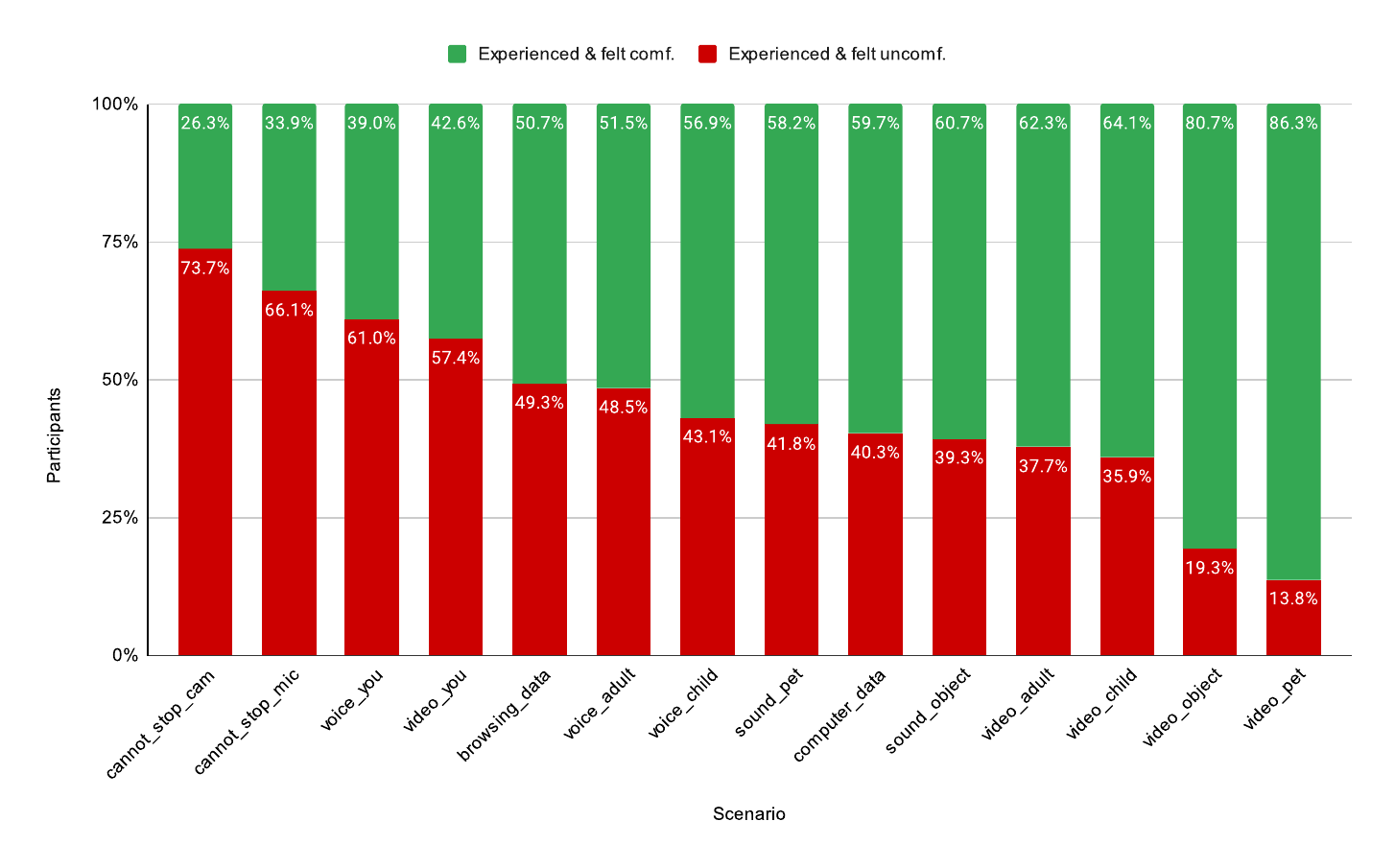}
\caption{Percentages of participants who experienced the named scenario and felt uncomfortable with respect to the total number of participants who experienced the named scenario.}
\label{fig:ratio_discomfort_prevalence_dissected}
\end{figure*}

\end{document}